\documentclass[12pt]{article}
\usepackage{graphicx}
\usepackage{amsfonts}
\usepackage{latexsym}
\usepackage{amsmath}

\makeatletter
\@addtoreset{figure}{section}
\def\thefigure{\thesection.\@arabic\c@figure}
\@addtoreset{table}{section}
\def\thetable{\thesection.\@arabic\c@table}

\def\@sect#1#2#3#4#5#6[#7]#8{\ifnum #2>\c@secnumdepth
     \def\@svsec{}\else
     \refstepcounter{#1}\edef\@svsec{\csname the#1\endcsname.\hskip .75em
}\fi
     \@tempskipa #5\relax
      \ifdim \@tempskipa>\z@
        \begingroup #6\relax
          \@hangfrom{\hskip #3\relax\@svsec}{\interlinepenalty \@M #8\par}%
        \endgroup
       \csname #1mark\endcsname{#7}\addcontentsline
         {toc}{#1}{\ifnum #2>\c@secnumdepth \else
                      \protect\numberline{\csname the#1\endcsname}\fi
                    #7}\else
        \def\@svsechd{#6\hskip #3\@svsec #8\csname #1mark\endcsname
                      {#7}\addcontentsline
                           {toc}{#1}{\ifnum #2>\c@secnumdepth \else
                             \protect\numberline{\csname the#1\endcsname}\fi
                       #7}}\fi
     \@xsect{#5}}
\def\@begintheorem#1#2{\it \trivlist \item[\hskip \labelsep{\bf #1\ #2.}]}
\def\section{\@startsection {section}{1}{\z@}{-3.5ex plus -1ex minus
 -.2ex}{2.3ex plus .2ex}{\normalsize\bf}}

\begin{document}

\title{How to Simulate Billiards and Similar Systems}
\date{} 
\maketitle

\begin{center}
\author{Boris D. Lubachevsky\\
{\em bdl@bell-labs.com}\\
Bell Laboratories\\
600 Mountain Avenue\\
Murray Hill, New Jersey}
\end{center}

\setlength{\baselineskip}{0.995\baselineskip}
\normalsize
\vspace{0.5\baselineskip}
\vspace{1.5\baselineskip}

\begin{abstract}
An $N$-component continuous-time 
dynamic system is considered whose components evolve autonomously
all the time except for in discrete asynchronous instances
of pairwise interactions.
Examples include
chaotically colliding billiard balls and
combat models. 
A new efficient serial event-driven algorithm is described
for simulating such systems.
Rather than maintaining and updating the global state of the system,
the algorithm tries to examine only essential events,
i.e., component interactions.
The events are processed in a non-decreasing order of time;
new interactions are scheduled on the basis
of the examined interactions using preintegrated
equations of the evolutions of the components.
If the components are distributed 
uniformly enough in the evolution space, so that this space
can be subdivided into small sectors such that only $O(1)$
sectors and $O(1)$ components are in the neighborhood of a sector,
then the algorithm spends time $O ( \log N)$ for processing
an event which is the asymptotical minimum.
The algorithm uses a simple strategy for handling data:
only two states are maintained for each simulated component.
Fast data access in this strategy assures the practical efficiency of
the algorithm.
It works noticeably faster than other algorithms
proposed for this model.
\\
\\
{\em Key phrases:} 
\\
collision detection, dense packing, molecular dynamics,
hard spheres, granular flow
\end{abstract}
\section{Introduction}\label{sec:intro}
\hspace*{\parindent} 
Many continuous time dynamic systems 
can be accurately approximated by models 
whose components evolve autonomously
all the time except for in discrete asynchronous instances
of pairwise interactions.
A typical example is a set of chaotically colliding billiard balls.
Each ball moves along a straight line until
it collides with another ball or an immobile obstacle.
Only pairwise ball collisions are considered,
since the probability is zero 
that more than two balls are involved in the same collision.

Such ``billiards'' or ``hard sphere'' models have been in use
among computational physicists since the pioneering work
\cite{ALDER}.
Recently these models have attracted
the attention of simulationists \cite{HONT}, \cite{PAS}.
The task of simulation of such a model
is reconstruction of the history of each component.
Many models,
even as far from billiards as models of combat \cite{WIEL},
are conceptually similar to billiards.
The similarity is in the
techniques for handling spatial combinatorics
of multitude of asynchronous pairwise interactions.
Processing an interaction
(two-ball collision) or 
an autonomous evolution of a component
(moving a billiard ball along a straight line)
depends on the specific model in hand.

Most of the recent attention 
has been drawn to the 
parallelization of such a simulation \cite{HONT}, \cite{LSCS}.
However, it remains
not obvious how to write
a practically efficient serial algorithm
for the billiard-type simulation.
A ``naive'' serial algorithm advances
the global state of the billiards from collision to collision.
The states of all $N$ balls are examined and updated
at times $t_0 \leq t_1  \leq  t_2 \leq...$,
where $t_0$ is the initialization time
and $t_{i+1}$ is the nearest next collision time
seen at time $t_i$.
The naive scheme is inefficient for large $N$
because of high costs it incurs while performing the following actions:

(a) the same collision is repeatedly scheduled
an order of $N$ times until it occurs,

(b) at a typical cycle, most balls
are not participating in collisions;
still, they are examined by the algorithm.

Aside from costly actions (a) and (b),
there exists problem (c) of finding an inexpensive method
of determining the nearest collision for a chosen ball.
A straight-forward method is to compare the chosen ball
with $N-1$ others.
The standard improvement in this method is the division
of the pool table into an order of $N$ sectors.
Only balls in the neighboring sectors have to be checked
to determine the immediate collision
which reduces the work from $O(N)$ to $O(1)$
per one collision scheduled.

A natural idea for improvement in (a) and (b)
is to postpone examining and updating the state of a ball
until its collision.
Implementing this idea does not appear as easy as it might seem.
As the simulation progresses,
a scheduled collision of a given ball may require rescheduling.
The need for such rescheduling and the desire not to loose
information about already planned collisions
lead in \cite{ALDER} to
a complicated data structure and update scheme
called ``time-table'' in \cite{ERPW}.
Observe that, with all its computing inefficiency,
the naive scheme has an attractively simple
double-buffering data structure.
The structure consists
of only two copies of the global state vector,
the old and the new,
so that the new vector is computed on the basis of the old one
and, in turn, becomes the old one during the next cycle.

We propose a new serial algorithm for the simulations like billiards.
The attraction of this algorithm is that it utilizes
a simple and easy to handle double-buffering data structure,
while avoiding costly actions (a) and (b).
Problem (c) is handled in our algorithm
using the standard technique of sectoring.
In most cases the algorithm examines and processes only 
the events whose processing is unavoidable,
e.g., ball collisions and boundary crossings.
Sometimes, like the naive algorithm,
it also processes events
whose examining is not necessary.
However, the fraction of such overhead events is 
less than 15\% in most experiments, 
and does not grow with $N$,
while the speed-up due to
simplicity of data handling is substantial.
The proposed algorithm achieves the same theoretical
optimal performances as other published algorithms,
i.e., $O( \log N )$ instructions per one processed event
with sectoring and $O( N )$ without.
But its practical algorithmic 
(i.e., computer independent)
speed for the billiard case
is at least an order of magnitude 
higher than that of other algorithms.
We were able to handle thousands balls
and millions of collisions on a non-supercomputer
VAX8550 using FORTRAN
(using languages better adapted for the computer
should result in additional speed-ups).
Figure~\ref{fig:2000ball} in Section~\ref{sec:pack} 
represents results of the experiments
for 2000 balls.
Using the same program on the same computer,
the number could be easily increased to $10^4$.

When writing this algorithm we paid special attention
to an often overlooked tradeoff
between the complexity of data organization
and the amount of computations
the algorithm is willing to abandon
risking to incur them again.
For example, to determine the next collision
for a ball $A$ we have to try to collide it
with any other ball 
(within its neighborhood, in the presence of sectoring)
and then choose the collision closest in time,
say it is a collision with ball $B$.
However this collision of $A$ and $B$ 
may not necessarily be the one which would
take place because the tentative partner $B$,
in its turn, may 
choose an even better party $C$, $C \neq A$.
As a result, 
later in the computations,
$A$ might 
figure out that the party $B1$,
which was previously rated second to best,
is to be considered the best.

Rating potential candidates costs computations:
the algorithm tries to simulate a potential collision of ball $A$
with a candidate $X$ in order to rate this $X$.
Should the algorithm retain the results of these preliminary simulations
which correspond to the second, third etc. to best parties $B1$, $B2$,... 
when the best candidate $B$ is being chosen ?
Or is it more economical to abandon the information
obtained during the rating
and, if later needed, simulate these collisions again ?
The answer determines the data organization strategy
which crucially affects algorithm efficiency.
In the billiard case,
a "pack rat" strategy entailing a search 
through dumped items to find the needed one incurs
too high a cost.
In a general case,
the best tradeoff 
depends on the relative
cost of the basic operations,
e.g., the amount of computing needed 
for repeated scheduling
versus that needed to retrieve the same data.

Another scale of strategies and the associated tradeoff
is that of the ``aggressiveness''
of precomputation.
In a more aggressive strategy,
when the next collision for ball $A$ is being
scheduled, not only the existing states of other
balls are taken into account but also
their possible future states which might
result from their as yet unprocessed collisions.
The degree of aggressiveness might be measured
in how many future collisions the other balls are being looked ahead.
The two scales are not independent:
a more aggressive precomputation
requires a more complicated data structure
and encourages the choice of 
a more ``pack rat'' data handling strategy.

In the described two tradeoff scales,
the strategy used in the proposed algorithm is close to
the ``wasteful'' and ``lazy'' ends  of the scales,
the opposite of the ``pack rat'' and ``aggressive'' ones.
Both the storage of not immediately needed data
and precomputation lookahead are reduced.
The candidates for the next collision for a particular ball 
which are rated below the winner 
are abandoned after the winner is chosen.
The future collision of a ball is predicted 
based only on the existing states of the other balls, 
not on their future states
after possible future collisions.

For a reader who is not familiar 
with simulation terminology,
it is worth adding that the proposed is
an {\em event-driven} simulation algorithm
in which the state of the simulated system
is examined by the computer only at the times of
the {\em events}, e.g., ball collisions.
A computational physicist may be more familiar
with the {\em time-driven} simulation algorithms.
Such algorithms (see, e.g., \cite{KATZ})
are usually employed in the many-body problems
in which the components, say particles, rather than evolving 
almost always autonomously,
are continuously interacting by exerting short and/or
long range forces.
The two computational approaches are radically different
and each has its own difficulties.

A time-driven algorithm would maintain
the snapshot of the states of all the simulated components
at a time $t$ and would advance the time from
$t$ to $t+ \Delta t$ by modifying all these states.
Given the same precision of simulation,
$\Delta t$ should be rather small
for billiards.
As a result,
the time-driven algorithm
would be tremendously slower 
than the proposed event-driven algorithm.
Event-driven algorithms are the best 
(and often the only practical) 
choice for models where 
discrete instantaneous events occur asynchronously.
In the proposed algorithm,
if at time $t$ an event involving ball $A$
is processed, 
only the state of $A$ is examined and
explicitly modified.
The states of most other balls
need not be known at $t$ and
are not examined by the algorithm.
In fact, the global state is explicitly known
at no time $t$,
except $t = 0$.
However, if 
we wish to know the global state,
say, if we wish to know the location of each ball 
at a particular time $t$,
then additional computations of
``projecting'' the motions of all balls
into time point $t$ are required.

The rest of the paper is organized as follows:
In Sections~\ref{sec:basic} to \ref{sec:comm},
a definition of the basic operations, the data organization,
the formulation of the algorithm with examples of its run,
and some comments on the practical experience of its implementation
are given.
These sections should be sufficient
for a reader who wishes to understand and write a  
simulation algorithm for a billiard-like system.
Sections~\ref{sec:consist} and \ref{sec:proof}
introduce,  explain, and analyze 
the conditions under which this algorithm
works correctly.
Section~\ref{sec:pack} presents an application example
for the algorithm: a disk packing problem,
Section~\ref{sec:perf} compares the performance of this algorithm
with other published proposals,
and Section~\ref{sec:other} discusses variants of the billiard simulation 
and other simulation models like billiards
including combat models.

\section{Basic operations}\label{sec:basic}
\hspace*{\parindent} 
Assume that a basic function $interaction\_time$
is available which,
given $state1$ of component 1 at $time1$
and
$state2$ of component 2 at $time2$,
computes the $time$ of the next potential interaction
while ignoring the presence of other system components:
\begin{equation}
\label{inttime}
time  \leftarrow  interaction\_time ( state1, time1, state2, time2)
\end{equation}
where $time \geq \max (time1, time2)$.
If $interaction\_time$ can not find
such finite $time$,
e.g., when two billiard balls are moving away from each other,
we assume that $+ \infty$ is returned.

\begin{figure}
\centering
\includegraphics*[width=4.8in]{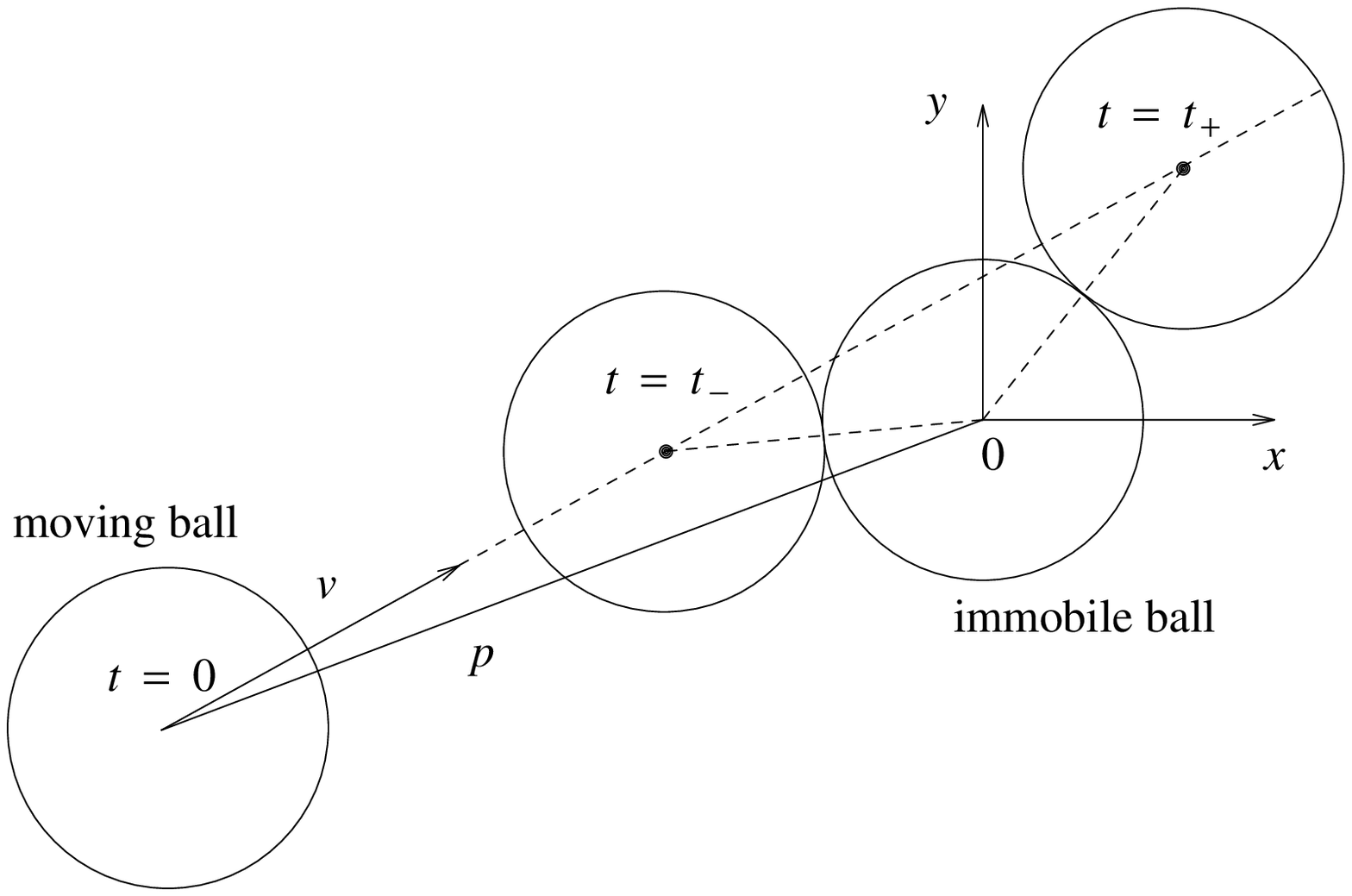}
\caption{The geometrical meaning of the two solutions of 
equation $ | p + v t |^2 = D^2$}
\label{fig:geomean}
\end{figure}

In the billiard simulation,
the state of a ball
is the pair of vectors $state = (position, velocity)$.
If the velocities of the balls are constant between the collisions,
and all balls are of the same constant diameter $D$,
then
\eqref{inttime} is of the form
\[
time  \leftarrow \max (time1, time2) + t
\]
where
\begin{equation}
\label{root1}
t = \left\{ \begin{array}{ll} 
(-b- \sqrt{b^2 - ac} )/a, & \mbox{if $b \leq 0$ and $b^2 -ac \geq 0$} \\
 + \infty ,                & \mbox{if $b > 0$ or   $~b^2 -ac < 0$}
\end{array}
\right. 
\end{equation}
and
\[ a = | velocity2 - velocity1 |^2 ,~~~~~~~~~~~~~~~~~~~~~~~~~~~~~~~~ 
~~~~~~~~~~~~~~~~
\]
\[ b = \langle position20 - position10, velocity2 - velocity1 \rangle , 
~~~~~~~~~~~~~~~~
\]
\[ c = | position20 - position10 |^2 - D^2 ,~~~~~~~~~~~~~~~~~~~~ 
~~~~~~~~~~~~~~~~
\]
\[position10 = position1 + velocity1 ( \max (time1, time2) - time1),\]
\[position20 = position2 + velocity2 ( \max (time1, time2) - time2),\]
$| v |$ denotes the length of vector $v$ and
$\langle u,v \rangle$ denotes the dot product of vectors $u$ and $v$.
The expression for $t$ in \eqref{root1} 
is the least real solution $t = t_-$
of the 
equation $a t^2 + 2 b t + c = 0$ which
is derived from $| p + v t |^2  = D^2$
where $p = position20 - position10$
and $v = velocity2 - velocity1$.
The meaning of the latter equation and of both its solutions 
$t = t_-$ and $t = t_+$ is
obvious from Figure~\ref{fig:geomean}.
Note that $c$ may equal 0 in which case $t = t_- = 0$
and $time =  \max ( time1, time2 )$.
This means that 
$interaction\_time$ is applied when one ball
is already at the site of the scheduled collision.

Two components 1 and 2 with $state1$ and $state2$ 
are said to be {\em interacting},
if
\begin{equation}
\label{already}
interaction\_time (state1, time, state2, time) = time
\end{equation}
holds for any value of $time$.
For example, billiard balls $i$ and $j$ of diameter $D$ each
with velocities and positions
$v_i , p_i$ and $v_j , p_j$, respectively,
are interacting (i.e., colliding),
if
\begin{equation}
\label{colliding}
|p_j  - p_i | = D, \langle v_j - v_i , p_j  - p_i \rangle \leq 0
\end{equation}

\begin{figure}
\centering
\includegraphics*[width=4.4in]{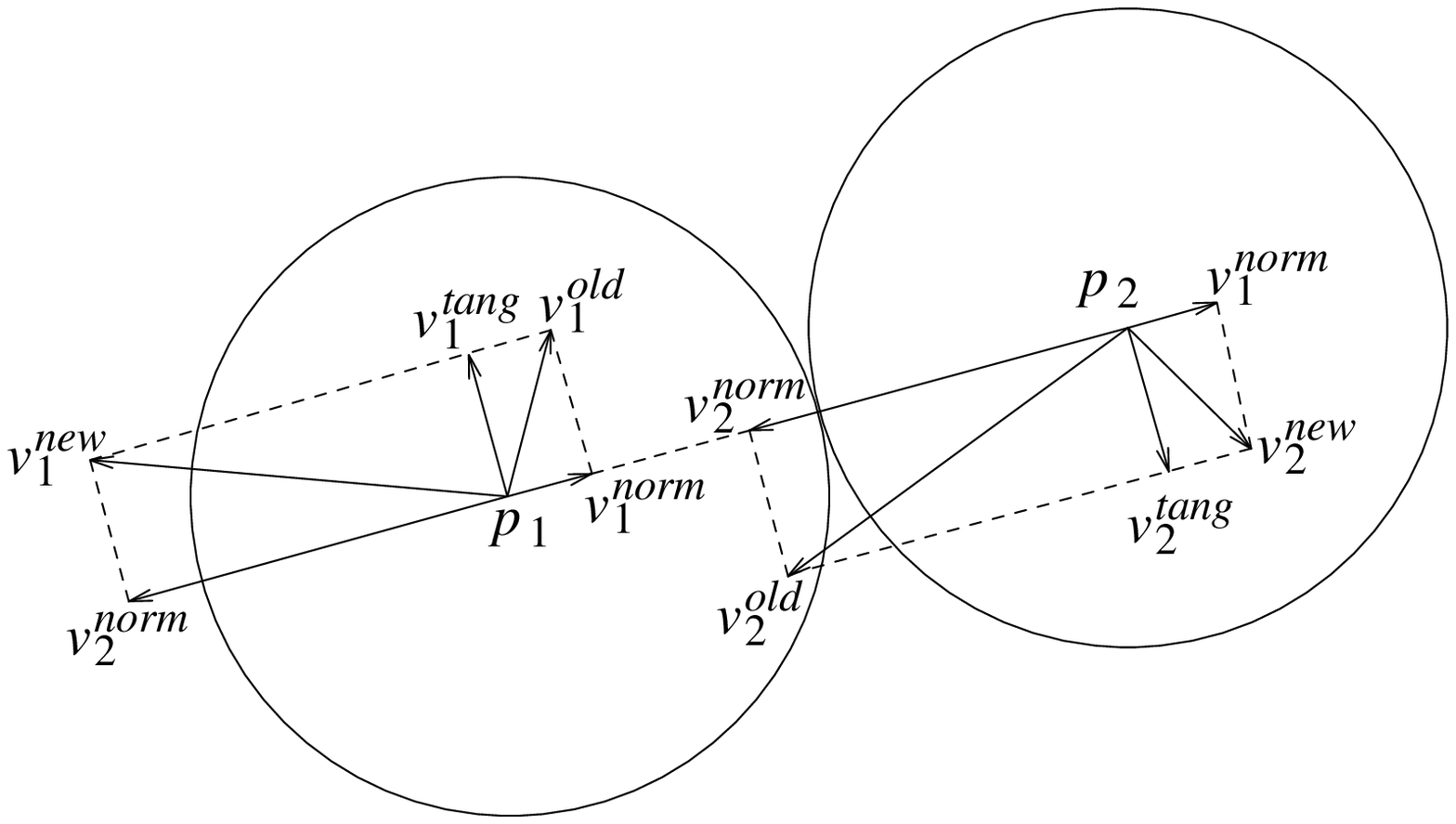}
\caption{Change of velocities of two billiard balls at their collision}
\label{fig:collis}
\end{figure}

Assume that a basic function $jump$
is available which,
given $state1$ and $state2$ of interacting components 1 and 2,
computes $new\_state1$ and $new\_state2$ of these components
immediately after the interaction:
\begin{equation}
\label{jump}
( new\_state1, new\_state2) \leftarrow jump (state1, state2)
\end{equation}
When two billiard balls collide, 
only their velocities 
experience jumps,
not positions.
Assuming the energy and momentum are conserved,
the tangential components of the initial velocities
are not changed, but the normal coordinates are switched
as depicted in Figure~\ref{fig:collis}.

A ball bouncing off a boundary of the pool table,
in principle, needs not be examined by the algorithm.
It may be considered as
an ordinary point on the autonomous
interval of the trajectory.
For example, given positions $( x0 , y0 )$
and the velocity vector $v$ of a ball at time $t = 0$,
we can construct functions $X( v , x0 , y0 , t )$ and $Y( v , x0 , y0 , t )$
which would return the position $x = X$ and $y = Y$
of the ball at time $t$ without explicitly processing
intermediate boundary reflections.
The complexity of computations by functions $X( )$ and $Y( )$
would not depend on the number of bouncing.

In most applications, however, such elaboration
would be of little practical use, because 
the ball would usually collide with another
ball after at most one boundary reflection.
Besides,
an explicit examination of the reflection event might be needed
anyway for statistical purposes and for convenience of data update.
Thus we will treat a 
reflection from
an immobile obstacle as a separate event.

\begin{figure}
\centering
\includegraphics*[width=2.8in]{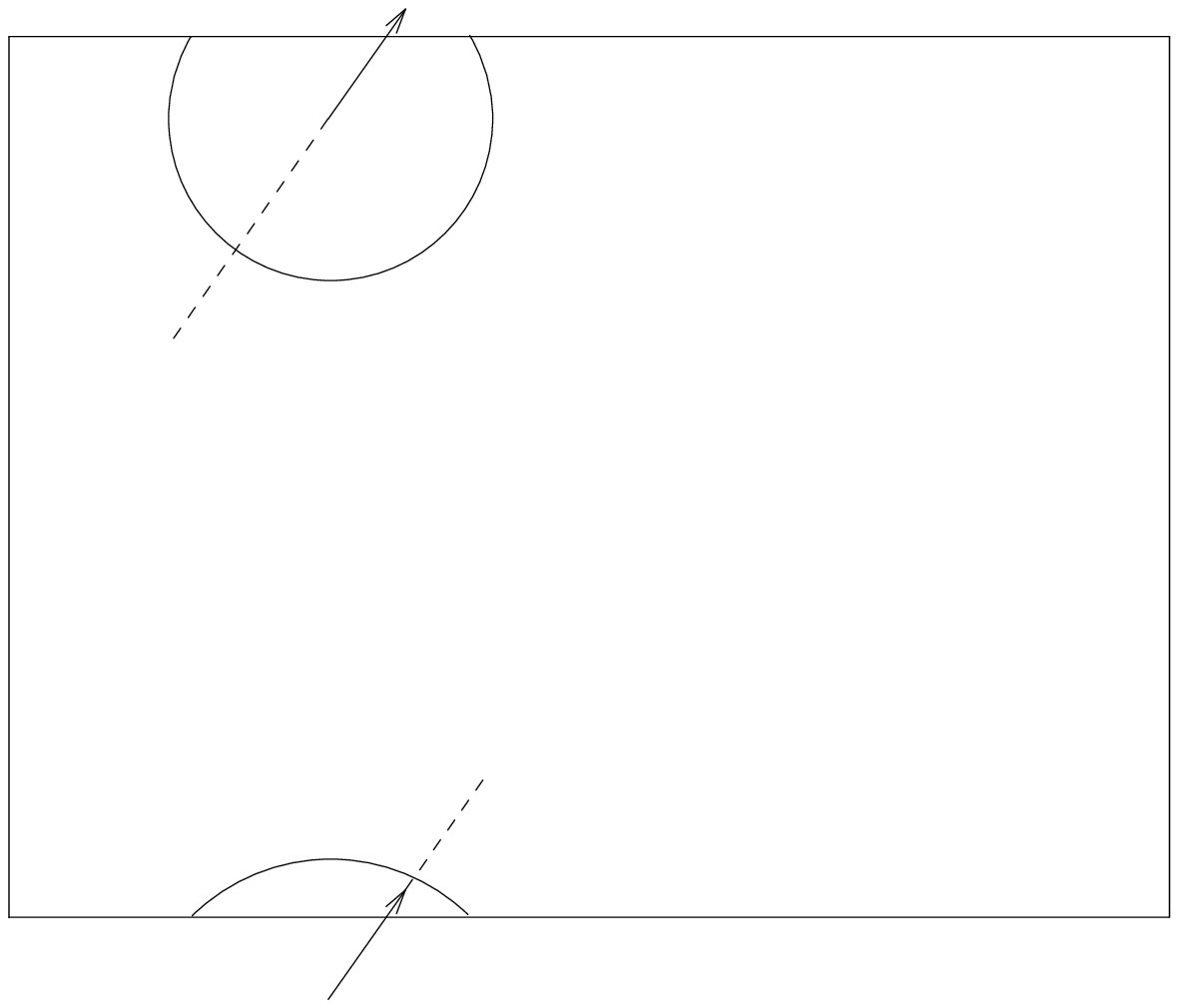}
\caption{A ball is disappearing at a boundary 
and reappearing at the opposite side
under periodic boundary conditions}
\label{fig:disapp}
\end{figure}

A boundary crossing
may be considered under a periodic boundary condition model,
wherein a ball,
rather than bouncing off,
disappears at a boundary and reappears at the opposite side
(see Figure~\ref{fig:disapp}).
This may be treated as the same type of event
as boundary reflection.
If the pool table is divided into sectors,
a similar type of event constitutes a ball
moving from one sector to another.
Such an event should
be examined by the algorithm in order
to update the membership in the sectors.
We will treat all such events as one-component interactions.

We will assume that basic
functions with the same names $interaction\_time$ and $jump$
represent
one-component interactions
\begin{equation}
\label{int0tim}
time  \leftarrow interaction\_time ( state1, time1, obstacle)
\end{equation}
\begin{equation}
\label{j0ump}
new\_state \leftarrow jump (state, obstacle)
\end{equation}
where $obstacle$ is the identification
of a boundary or an immobile obstacle or a demarcation line.
To apply $jump$ in \eqref{j0ump}, the component, whose $state$ 
is represented in \eqref{j0ump}, must be {\em interacting}
with $obstacle$.
The condition which defines such one-component interaction
is:
\begin{equation}
\label{alre0dy}
interaction\_time (state1, time, obstacle) = time
\end{equation}
holds for any $time$.
This is similar to \eqref{already}.
Capital $K$ will be reserved for
the number of obstacles
so that $obstacle$ in 
\eqref{int0tim}, \eqref{j0ump}, and \eqref{alre0dy} is an integer
in the interval from 1 to $K$.
The one-component versions of functions 
$interaction\_time$ and $jump$ will
be easily distinguished by context
from their two-component synonyms in \eqref{inttime} and \eqref{already}.

We also assume the availability of a basic function $advance$ which,
given $state0$ of a component at $time0$ and a value $time1 \geq time0$,
returns $state1$ this component would have at $time1$ 
ignoring possible interactions with other components
or obstacles
on the interval $(time0, time1)$:
\begin{equation}
\label{adva}
state1 \leftarrow advance ( state0, time0, time1)
\end{equation}
In a frictionless billiard,
\eqref{adva} is of the form
\begin{equation}
\label{mov}
\begin{array}{l}
 position1 \leftarrow position0 + (time1 - time0 ) velocity0 \\
 velocity1 \leftarrow velocity0
\end{array}
\end{equation}
which simply says that the ball
moves with $velocity0$ along a straight line 
starting from $position0$ at $time0$.

Note that in particular cases,
specific calculations for $interaction\_time$,
$advance$, and $jump$ 
may be not as simple as in the billiard case.
The assumption that they are {\em basic}
saves us from the burden to detail them in the general discussion.

\section{Data organization}\label{sec:data}
The basic data unit of the algorithm
is called {\em event} and has
the following format
\begin{equation}
\label{event}
event  =  ( time, state, partner)
\end{equation}
where:

$time$ is the time to which $state$ of a component corresponds.
Note that $state$ is the new state of the component 
{\em immediately 
after} the event,
e.g., if a ball has experienced a collision at $time$,
the velocity-coordinate of the $state$ is the new velocity vector
after the collision;

$partner$ identifies the other component, if any, involved in the event.
If there is no partner in the event,
the program assigns a special ``no-value" symbol
$\Lambda$ to the $partner$ coordinate.

If $time = + \infty$,
then the other three coordinates in the $event$ have no value,
i.e., $state=type=partner=\Lambda$.

At any stage of simulation,
the algorithm maintains two events for each component:
an old, already processed in the past event
and a new, next scheduled event.
This information is
stored in array $event[1:N,1:2]$,
where $N$ is the number of components of the simulated system.
Let us agree to understand a reference
like $time[3,1]$ as the $time$ coordinate of element $event[3,1]$
of this array.

Two arrays $new[1:N]$ and $old[1:N]$ with elements equal 1 or 2 are maintained.
The value $new[i]$ is the pointer to the new event for component $i$
and
the value $old[i]$ is the pointer to the old event for component $i$,
so that new event for component $i$ is stored at
$event[i,new[i]]$ and old event for component $i$ is stored at 
$event[i,old[i]]$.
When $new[i]$ is updated, $old[i]$ is updated immediately after,
so that relation $new[i] + old[i] = 3$ remains invariant.

\section{The algorithm}\label{sec:algo}
\hspace*{\parindent} 
In the algorithm pseudocode in Figure~\ref{fig:algo1}, 
/" and "/ mark the beginning and the end of a comment,
the minimum over an empty set of values is assumed $+ \infty$,
and the following short-hand notations are used:
\\
\[ 
P_{ij} 
\stackrel{\rm def}{=} 
interaction\_time 
(state[i,old[i]], time[i,old[i]], state[j,old[j]], time[j,old[j]]) .
\]
\\
where $1 \leq i, j \leq N$ and 
\\
\[
Q_{ik} 
\stackrel{\rm def}{=} 
interaction\_time (state[i,old[i]], time[i,old[i]], k)
\]
\\
where $1 \leq i \leq N$ and $1 \leq k \leq K$.

The main cycle in Figure~\ref{fig:algo1}
essentially consists of two steps:
\\
1) selecting the next component $i_*$ to process its event (line 2),
\\
2) processing the event (the rest of the cycle).

Processing of the event comprises 
scheduling of the next events for the chosen component
and for the other components involved, if any.
{\bf P} and {\bf Q} are the nearest next interaction times.
There are two main cases in such scheduling
depending on the type of the future event:
  
a) when {\bf Q} $<$ {\bf P}, 
scheduling an interaction 
which involves only the chosen component $i_*$
(lines 8, 10, and 11);
  
b) when {\bf Q} $\geq$ {\bf P}, 
scheduling an interaction
which involves also a second party $j_*$
(lines 8 and 13 - 17)
and may involve a third party $m_*$,
the previous partner, if any, of $j_*$
(lines 19 and 20).

Section~\ref{sec:example} further explains
this simple algorithmic structure 
in 
concrete examples of its execution.
Now we discuss the aspects of the algorithm
which are {\em not} represented in the aggregated pseudocode
in Figure~\ref{fig:algo1},
namely, the ways the three minimizations in lines 2, 4, and 5 are implemented.
Since these techniques are well-known, their discussion will be brief.

Note that for a small number of balls, say $N = 10$, the best way
to minimize is to use the straight-forward element-by-element testing.
Such straight-forward method to find the minimum 
of $time[i,new[i]]$ for $i$ ranging from 1 to $N$
in line 2 requires $O(N)$ operations per event.
Instead, the algorithm organizes values $time[i,new[i]]$
into an implicit {\em heap} structure
with two pointer arrays $pht[1:N]$ and $pth[1:N]$
so that $time [pht[m], new[pht[m]]]$ is the 
value which is implicitly located at the $m$-th 
position of the imaginary heap array
and $pth$ is the inverse map for $pht$,
i.e., $pth [pht [m]] = m$ for all $m$.
In particular,
$time [pht[1], new[pht[1]]]$ corresponds
to the heap tree root, i.e., the minimum value.
Thus line 2 could be simply rewritten as
\\
\[
i_* \leftarrow  pht[1], ~~~~
current\_time \leftarrow time[i_* , new[i_* ]]
\]
\\
The payment for this computationally cheap method
of finding the minimum is the need to 
update the heap structure,
i.e., arrays $pht$ and $pth$,
each time a value of $time[i,new[i]]$ is changed
in other sections of the algorithm.
The total cost of finding the minimum next event time
thus runs to as much as $O( \log N )$ operations per one event.
For a large $N$,
the cost $O( \log N )$ is still much better than the original cost of $O(N)$.

The main difficulty in the straight-forward method
for minimizations in lines 4 and 5
is the need for computing the $N-1$ values 
$P_{i_* j}$ in line 4 and
the $K$ values  $Q_{i_* k}$ in line 5.
The opportunity to decrease the $O(N+K)$ complexity burden 
of these computations
depend on the topology of the evolution space
and the uniformity of the component and
obstacle distribution
in this space.
In the billiard case
the space is just the Euclidean plane
and there is an upper bound on the number
of balls which can be located in a bounded vicinity of a given ball.
To take a computational advantage of this,
the simulation space is divided into sectors
and only the components
or obstacles incident to the neighboring sectors
are examined.
The sector boundaries naturally become additional
``obstacles'' in this method
and their processing constitute the method's overhead.
However the practical gain in the method is high as examples
in Section~\ref{sec:example} and \ref{sec:pack} show.
Theoretically, the cost reduces to $O(1)$ when using this method.

Among the available grids used for planar sectorization,
the grid of equal squares is the most convenient.
(The ratio of the area to the perimeter 
of a sector is larger for the hexagonal grid,
though.)
Specifically in the case of equal balls,
we usually choose square sides larger than the ball diameter,
and for each square we maintain the membership linked-list of balls
whose centers project to this square.
Processing of a square boundary crossing is accompanied
by the update of the two lists.
Only those $P_{i_* j}$ are computed
and are subject to minimization in line 4,
for which the center of ball $j$ belongs
to one of the nine sectors neighboring
the one whose member is $i_*$.
\\
\begin{figure}
\centering
\fbox{
\fbox{
\begin{minipage} {15.1cm}
{\small
initially $current\_time \leftarrow  0$ and for $i = 1,2,...N$ :
$new[i] \leftarrow 1$, 
$old[i] \leftarrow 2$, 
$time[i,1] \leftarrow 0$, 
\\
$partner[i,1] \leftarrow \Lambda $,
$state[i,1] \leftarrow$ initial state of component $i$, 
$event[i,2] \leftarrow event[i,1]$
\\
---------------------------------------------------------------------------------------------------------------------
\begin{enumerate}
\item 
while $current\_time < end\_time$ do \{
\item~~~
$current\_time \leftarrow {\min}_{1 \leq i \leq N} time [i, new[i]]$ ; 
\\
$~~~~~~~~i_* \leftarrow$ an index which supplies this minimum 
(i.e., $current\_time$) ;
\item~~~
$new[i_*] \leftarrow old[i_*]$;   $old[i_*] \leftarrow 3-new[i_*]$ ;
\item~~~
{\bf P} $\leftarrow {\min}_{j \in A(i_*)} P_{i_* j} $,
where $A(i_*) = 
\{1 \leq j \leq N, j \neq i_*, time[j,new[j]] \geq P_{i_* j} \}$ ; 
\\
$~~~~~~~$
if {\bf P} $< + \infty$ then 
$j_* \leftarrow$ an index which supplies this minimum 
(i.e., {\bf P}) ;
\item~~~
{\bf Q} $\leftarrow {\min}_{k \in B} Q_{i_* k} $,
where $B =
\{1 \leq k \leq K\}$ ;
\\
$~~~~~~~$
if {\bf Q} $< + \infty$ then 
$k_* \leftarrow$ an index which supplies this minimum 
(i.e., {\bf Q}) ;
\item~~~
{\bf R} $\leftarrow \min $\{{\bf P}, {\bf Q}\};  
$time [i_*, new[i_*]] \leftarrow$ {\bf R};
\item~~~
if {\bf R} $< + \infty$ then \{
\item~~~~~~
$state1 \leftarrow 
advance (state[i_*, old[i_*]], time[i_*, old[i_*]]$, {\bf R} ) ;
\item~~~~~~
if {\bf Q} $<$ {\bf P} then \{
\item~~~~~~~~~
$state [i_*, new[i_*]] \leftarrow jump (state1,k_* )$ ;
\item~~~~~~~~~
$partner [i_*, new[i_*]] \leftarrow \Lambda$ ;
\\
$~~~~~~~$ \} /" end {\bf Q} $<$ {\bf P} clause "/ 
\item~~~~~~
else \{    /" case {\bf Q} $\geq$ {\bf P} "/
\item~~~~~~~~~
$time[j_*, new[j_*]] \leftarrow$ {\bf R} ;
\item~~~~~~~~~
$state2 \leftarrow 
advance (state[j_*, old[j_*]], time[j_*, old[j_*]]$,{\bf R}) ;
\item~~~~~~~~~
$(state [i_*,new[i_*]],state [j_*, new[j_*]]) \leftarrow jump(state1,state2)$ ;
\item~~~~~~~~~
$m_* \leftarrow partner [j_* , new[j_* ]]$ ;
\item~~~~~~~~~
$partner [i_* , new [i_* ]] \leftarrow j_*$ ; 
$partner [j_* , new [j_* ]] \leftarrow i_*$ ;
\item~~~~~~~~~
if $m_* \neq \Lambda$ and $m_* \neq i_*$ then \{ /" update third party $m_*$"/
\item~~~~~~~~~~~~
$state [m_* , new[m_* ]] \leftarrow $
\\
$~~~~~~~~~~~~~~~~~~$
$advance(state[m_*, old[m_*]], time[m_*, old[m_* ]], time[m_* , new[m_* ]] )$ ;
\item~~~~~~~~~~~~
$partner [m_* , new[m_* ]] \leftarrow \Lambda$ ;
\\
$~~~~~~~~~~~$\} /" end update third party "/ 
\\
$~~~~~~~~$\} /" end {\bf Q} $\geq$ {\bf P} clause "/
\\
$~~~~~$\}  /" end {\bf R} $< + \infty$ clause "/
\\
$~~$\}  /" end while loop "/
\end{enumerate}
}
\end{minipage}}}
\caption{
The simulation algorithm
}
\label{fig:algo1}
\end{figure}
\section{Simple example}\label{sec:example}
\begin{figure}
\centering
\includegraphics*[width=5.8in]{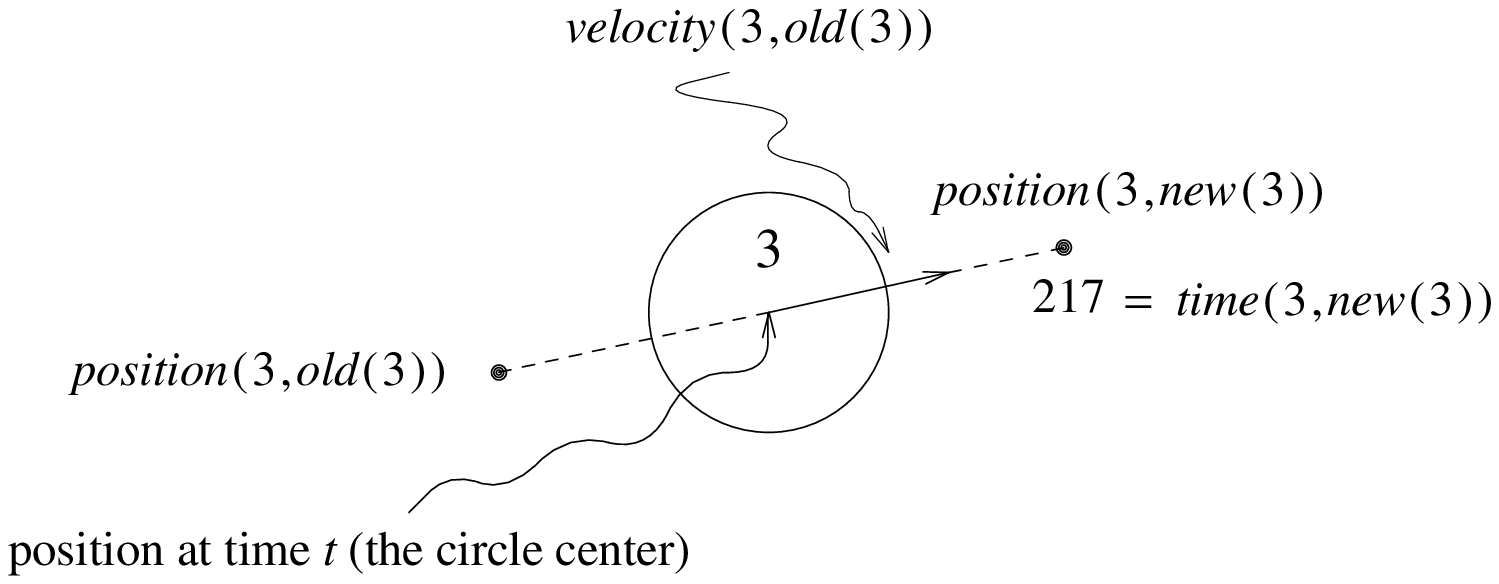}
\caption{Ball 3 at time $current\_time = t$ 
(a legend for Figure~\ref{fig:ex1} and \ref{fig:ex2})}
\label{fig:legend}
\end{figure}

\begin{figure}
\centering
\includegraphics*[width=6.3in]{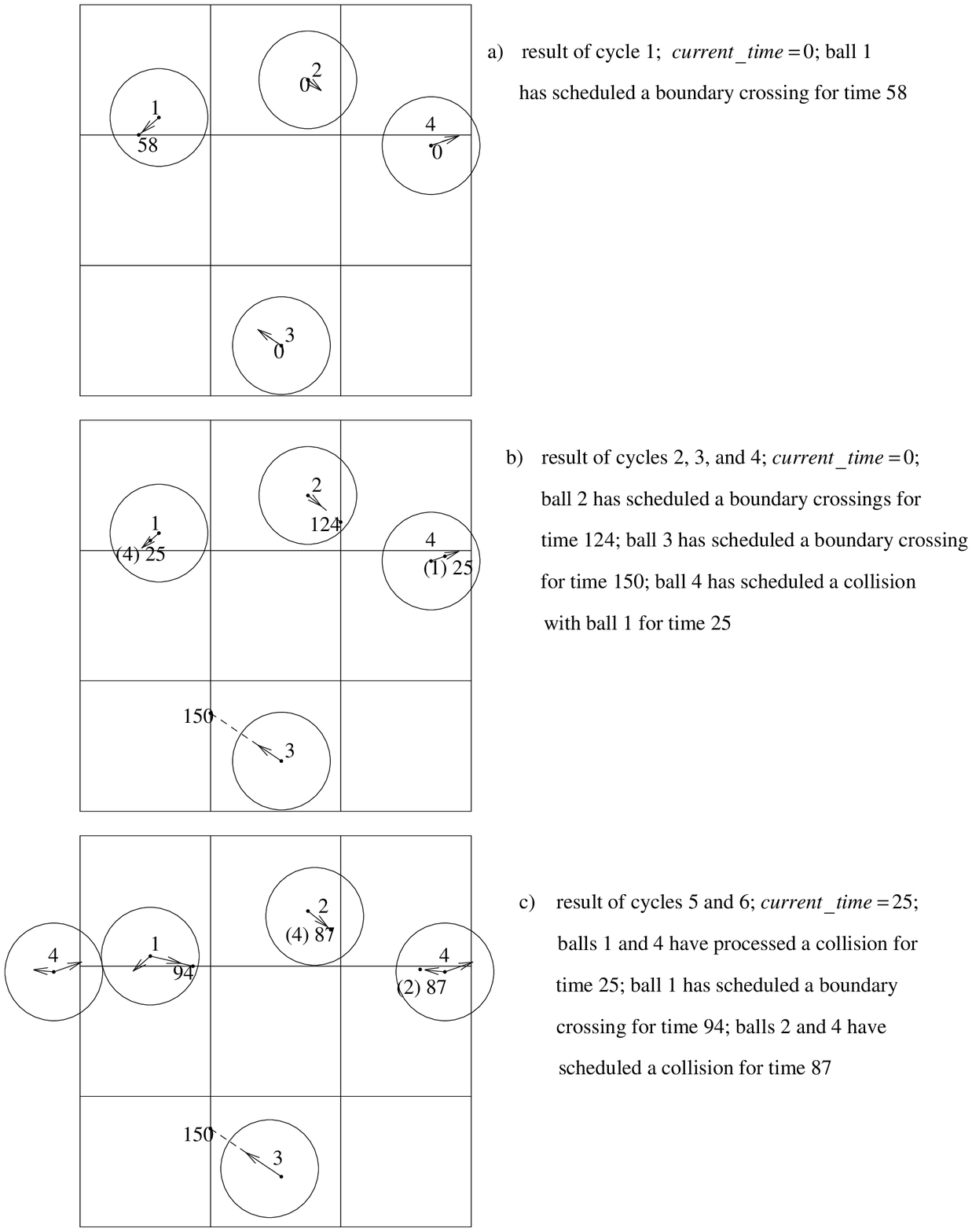}
\caption{An example of the algorithm execution. Space is sectorized}
\label{fig:ex1}
\end{figure}

\begin{figure}
\centering
\includegraphics*[width=6.1in]{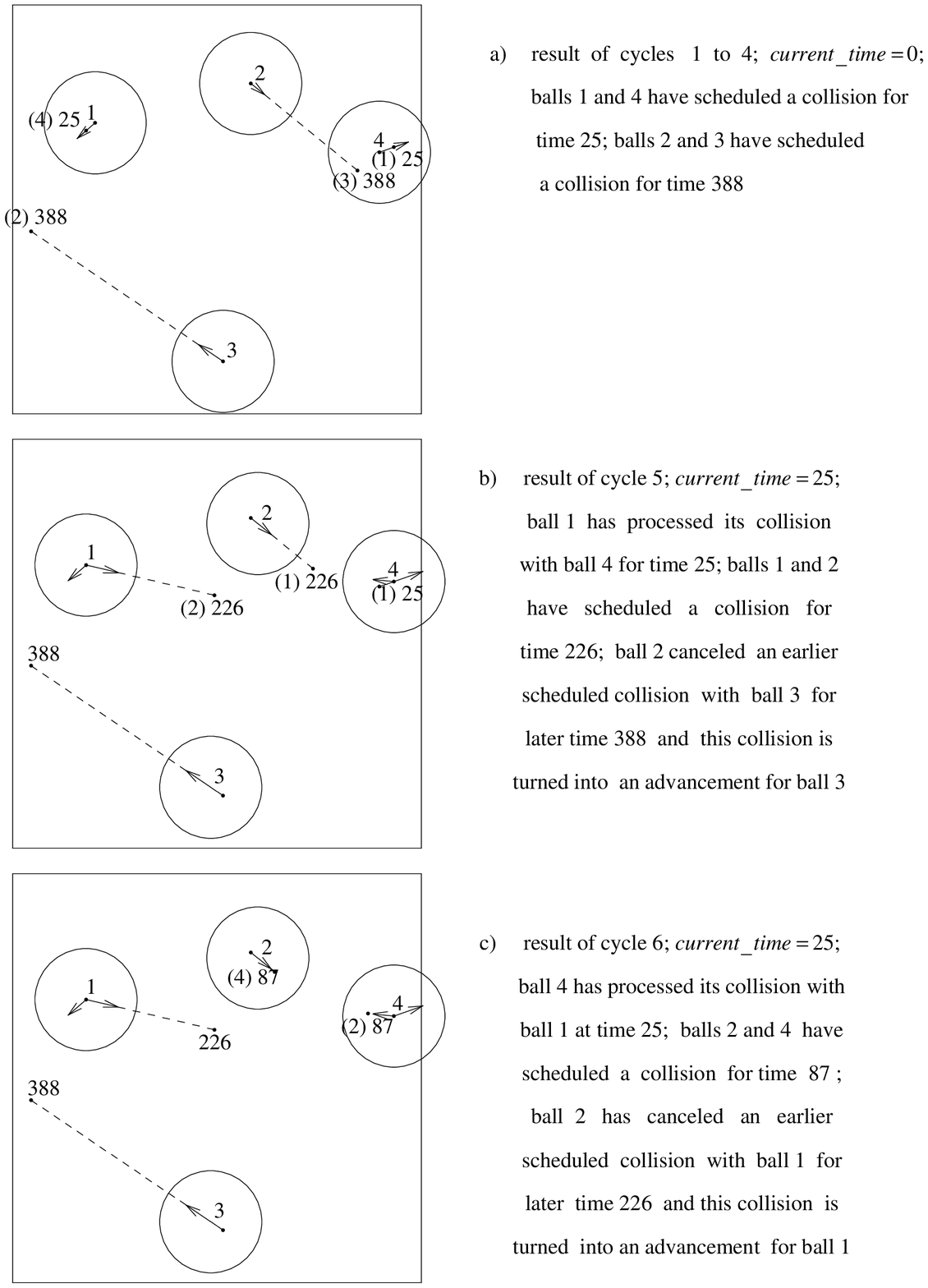}
\caption{An example of the algorithm execution. 
Space is not sectorized.
The initial conditions for the balls are the same as in Figure ~\ref{fig:ex1}}
\label{fig:ex2}
\end{figure}

\hspace*{\parindent} 
Two examples of the execution of
the algorithm in Figure~\ref{fig:algo1} are reproduced 
in Figures~\ref{fig:ex1} and \ref{fig:ex2}.
Four-ball billiards are simulated in both examples.
The pool table is a square.
Unlike real billiards with hard wall boundaries,
periodic boundary conditions are assumed
(these conditions are explained in Section~\ref{sec:basic}, 
see Figure~\ref{fig:disapp}).
In the example shown in Figure~\ref{fig:ex1},
the table is subdivided into $3 \times 3$ equal square sectors.
Figure~\ref{fig:ex1}
consists of three frames,
a, b, and c.
Each frame shows a snapshot 
of the simulation state at a particular $current\_time$ with
the identification, position, and velocity vector of each ball
at this time.
Since the execution state usually does not contain positions
of all the balls
at the same $current\_time$,
a picture-producing routine
(not presented in this discussion)
accepts $t = current\_time$ as an input  and
interpolates between the $old$ and the $new$
positions of the ball as shown in 
Figure~\ref{fig:legend}.
Note that while 
Figure~\ref{fig:legend}
shows a ``general'' case,
with $time[i,old[i]] < t < time[i,new[i]]$,
the snapshots in Figure~\ref{fig:ex1} and \ref{fig:ex2}
have many ``degenerated'' cases,
e.g., $time[i,old[i]] = t$.
Also note that for simplicity of the pictures,
the times are rounded off to their integer parts
(the computer manipulates them with the machine
precision for representing real numbers).

Figure~\ref{fig:ex1}a shows the positions and velocities
of the four balls at $current\_time = 0$.
These quantities are the initial values.
Observe that no two balls overlap.
(A method to define such initial positions 
is discussed in Section~\ref{sec:pack}.
Correct simulation 
should preserve this property.)
As the initialization statement in Figure~\ref{fig:algo1} reads,
the balls are initialized at the same zero $time$ with
identical $old$ and $new$ events.
Succeeding the test in line 1, Figure~\ref{fig:algo1} 
(assuming $end\_time$ is sufficiently large),
the algorithm is searching for a ball index $i_*$ which yields
the minimum to $time[i,new[i]]$.
As Figure~\ref{fig:ex1}a indicates, the algorithm has chosen ball 1.
Observe that in the beginning of simulation,
all four $new$ events have the same $time$ so 
the other three choices would be also correct.
After switching the senses of $old$ and $new$ event storages
for ball 1 in line 3
(here a redundant manipulation),
in line 4
the algorithm tries to select the ball
with which ball 1 will collide first.
Since $time[i, new[i]] = 0$ for all $i$ and all $P_{ij} > 0$,
no $j$ satisfies $time[j, new[j]] \geq P_{i_* j}$.
This means that the set subject to minimization 
in line 4 is empty.
Hence {\bf P}$ = + \infty$, and no $j_*$ is selected.
In line 5 
the algorithm selects the boundary $k_*$
which will be reached by ball 1 first.
This boundary happens to be
the lower side of the sector to which $position[1,old[1]]$ belongs
and the ball reaches it (in the absence of other balls)
at time {\bf Q} $= 58$.
In line 6, {\bf R} and $time[1,new[1]]$ are becoming equal to this time.
Tests in line 7 and 9 are succeeding and the rest of cycle 1
is spent on assigning to the $new$ coordinates their scheduled values
in lines 8, 10, and 11.
These new values will be in effect immediately after
crossing the specified boundary.
Note that if $obstacle$ is a demarcation boundary between sectors,
then $jump$ is defined as an identical function:
$jump (state, obstacle)
\stackrel{\rm def}{=} 
state$.
Then the algorithm takes the snapshot of the situation  
(and this snapshot is shown in Figure~\ref{fig:ex1}a),
after which cycle 2 is started.
On the snapshot, ball 1 has a scheduled event at time 58,
while the other three balls still have scheduled events
at time 0 as indicated.

Cycles 2, 3, and 4 are spent on scheduling the future events
with positive times
for the remaining three balls.
Figure~\ref{fig:ex1}b shows the progress
made in this scheduling.
While $current\_time$ is still at 0
because no event with positive time
has been processed yet,
balls 2 and 3 have scheduled boundary crossings
(case {\bf Q} $<$ {\bf P})
and ball 4 has scheduled
a collision at time 25 with ball 1
(case {\bf Q} $\geq$ {\bf P}).
When a scheduled collision is indicated on a picture,
not only its time is given but also (in parentheses)
the partner index. 
Thus, $(4)  25$ at the $new$ position of ball 1 means
that (the center of) ball 1 reaches this position
at time 25 and when it does so,
it collides with ball 4.
(The dashed line which is supposed to indicate the future
motion of ball 4 is overstricken by the arrow indicating the velocity.)

The algorithm schedules
this collision at cycle 3
when balls 1 and 2 have already scheduled their next events,
boundary crossings at times 58 and 124 respectively,
but ball 3 has not been touched by the algorithm yet.
This scheduling proceeds as follows.
First (line 4), ball $i_* = 4$ finds out
that the only $P_{4j}$ which is not larger than
$time [j,new[j]]$ is $P_{41} = 25$ and {\bf P} becomes 25.
Then (line 5), it is determined that the nearest boundary
crossing occurs at time {\bf Q}.
The smallest of the two, {\bf P} and {\bf Q}, 
becomes {\bf R} and also $time[4,new[4]]$ in line 6.
Since {\bf R} is finite and
{\bf Q} is larger than {\bf P},
test in line 7 succeeds but test in line 9 fails.
As a result, the sequence of statements in lines
8 and 13 - 17 is executed
whereby
balls 4 and 1 have scheduled a collision at time 25
and the index $m_*$ of the third party
is remembered.
Since there was no partner 
in the previously scheduled by ball 1 $new$ event,
$m_*$ becomes $\Lambda$ and lines
19 and 20 are skipped.

Time 25 becomes the smallest one in the event-list
and the next two cycles, 5 and 6, are spent
on processing two events,
$event[1,new[1]]$ and $event[4,new[4]]$,
both representing the collision of balls 1 and 4 at time 25
but from the ``viewpoints'' of two different balls.
Processing the collision event by ball 1 generates
new boundary crossing scheduled for time 94
and processing the collision by ball 4 generates another
collision scheduled for time 87 with ball 2.
The latter collision preempts the previously scheduled
by ball 2 boundary crossing for time 124.
The result of all these processings is shown in Figure~\ref{fig:ex1}c.
Two velocity vectors are indicated for each colliding ball
in Figure~\ref{fig:ex1}c: before and after the collision.
As seen, ball 1 has collided not with ball 4 but with its
periodic image.

The sequence of snapshots shown in Figure~\ref{fig:ex2}
corresponds to the same initial condition of the balls
as in Figure~\ref{fig:ex1}
but without sectoring.
During cycles 1 to 4 two collisions are scheduled:
one of balls 1 and 4 for time 25 and another of balls 2 and 3
for a distant time 388.
However after cycle 5 the more distant collision
of balls 2 and 3 is preempted by a collision
of balls 1 and 2 for earlier time 226.
As a result, ball 3 is left without a collision;
its tentative collision is turned into 
a no-partner event
which will be convenient to call {\em advancement}.
However, at cycle 6 the preemptor, collision of balls 1 and 2 for time 226,
is itself preempted by a
collision of balls 2 and 4 scheduled for even earlier time 87.
As a result, ball 1 has now scheduled an advancement event,
the one previously believed to be a collision scheduled for time 226.

It seems that events develop
faster in the experiments without sectoring shown in Figure~\ref{fig:ex2}
than in those with sectoring in Figure~\ref{fig:ex1}.
Without sectoring the balls attempt to schedule their new events with
larger horizons, they are more ``aggressive.''
However, each cycle here takes more computing time.
We have continued both experiments for $10^5$ collisions,
each pairwise collision counted twice.
Without sectoring, it takes more than three times longer of the CPU time,
than with sectoring.

This is so because
to schedule a collision with sectoring, 
a ball should check
nine neighboring sectors including its own,
where it finds at most three other balls.
Without sectoring a ball should check the same three balls,
and also their $3 \times 8$ periodic boundary images.
Functions $interaction\_time$ are formally different in the
two cases. 
In the case without sectors, time of a next collision 
with a ball $A$ is in fact given not as \eqref{root1}
but as the minimum of nine times,
one of which represents a collision with $A$
and is given by \eqref{root1} and the other eight
represent collisions with eight periodic images of $A$.

\section{Comments on the practical execution of the algorithm}\label{sec:comm}
\hspace*{\parindent} 
{\bf Overlaps}.
A practically working algorithm for billiard simulation
should be resistant with respect to small overlap of the balls.
Figure~\ref{fig:ex2} shows
``a preemption of a preemptor'' phenomenon
when ball 1 has preempted
a collision of balls 2 and 3 by scheduling an earlier
collision with ball 2 (Figure~\ref{fig:ex2}b)
only to be later preempted by
ball 4 which schedules an even earlier collision
with ball 2 (Figure~\ref{fig:ex2}c).
In the simulations with thousands
of balls, more involved phenomena of this kind 
occur.
Combined with the roundoff,
occasionally,
they cause small overlaps
of the balls
as shown in the following example.
Suppose, a scheduled collision of balls $A$ and $B$ 
for time $t_{AB}$
is later preempted by scheduling collision
of $B$ and $C$ for time $t_{BC} < t_{AB}$.
As a result, the collision event for $A$ becomes an advancement
for time $t_{AB}$.
Suppose, that later, 
collision of $C$ and $D$ scheduled for time
$t_{CD} < t_{BC}$ preempts the collision of $B$ and $C$.
As a result, 
the collision event for $B$ becomes an advancement
for time $t_{BC}$.
Now the originally scheduled collision of $A$ and $B$ for time $t_{AB}$
needs to be scheduled again. 
However, it will be done starting with different initial positions.
If formula \eqref{root1} is used in this scheduling,
then
$c = 0$ and $t = 0$ because $\max (time1, time2) = t_{AB}$.
Because of roundoff errors
and different computational paths,
$c$ may be ``slightly'' negative
as if balls $A$ and $B$ would slightly overlap at time $t_{AB}$
and so negative could be $t$.
In the existing program this problem is handled as follows:
whenever $interaction\_time$ computes a negative
but
small by absolute value $t$ in \eqref{root1},
this is not reported as an error,
but the value of $t$ is replaced by zero.

{\bf Advancement events}.
A preempted two-component interaction is turned
into an advancement for the third party,
e.g., the preempted collision for time 388 
of balls 2 and 3 in Figure~\ref{fig:ex2}a
is turned for ball 3 into an advancement in Figure~\ref{fig:ex2}b.
A more aggressive strategy
would perform a full-fledged new event scheduling
for ball 3.
Such strategy is less efficient partly because
advancements are usually far fetched in the future
and have a great chance to be rescheduled 
so it is not worthwhile to waste precomputations on them.
By the time of event processing
only a small fraction of events
remain advancements,
in most simulated cases less than 15\%.
More importantly, the fraction of the processed
advancements does not grow with $N$.
(No theoretical analysis of this statement is available.)
Another possibility is ``rolling back'' the preempted ball
to the $old$ event.
This would break the time-orderly 
event processing.

{\bf Delayed update}.
There exists a subtle
inefficiency in the algorithm in Figure~\ref{fig:algo1}.
When
scheduling an interaction, the algorithm applies
$advance$ and $jump$ operations.
If the event is later preempted,
these computations are wasted.
For example, when scheduling a collision
of balls 2 and 3 for time 388 (Figure~\ref{fig:ex2}a),
new velocities are computed, using $jump$.
Later, however, this collision is preempted
(Figure~\ref{fig:ex2}b).
To correct this inefficiency,
the application of $advance$ and $jump$
should be delayed until the latest possible moment
when the scheduled event is being processed.
Such optimized pseudocode 
(which looks less transparent than the original one) 
is given in Figure~\ref{fig:algo2}.
\\
\begin{figure}
\centering
\fbox{
\fbox{
\begin{minipage} {15.1cm}
\footnotesize{
initially $current\_time \leftarrow  0$ and for $i = 1,2,...N$ :
$new[i] \leftarrow 1$, 
$old[i] \leftarrow 2$, 
$time[i,1] \leftarrow 0$, 
\\
$partner[i,1] \leftarrow \Lambda $,
$state[i,1] \leftarrow$ initial state of component $i$, 
$event[i,2] \leftarrow event[i,1]$
\\
--------------------------------------------------------------------------------------------------------------------------------
\begin{enumerate}
\item 
while $current\_time < end\_time$ do \{
\item~~~
$current\_time \leftarrow {\min}_{1 \leq i \leq N} time [i, new[i]]$ ; 
\\
$~~~~~~~~i_* \leftarrow$ an index which supplies this minimum 
(i.e., $current\_time$) ;
\item~~~
$state1 \leftarrow 
advance (state[i_*, old[i_*]], time[i_*, old[i_*]],
time[i_*, new[i_*]])$ ;
\item~~~
$j_{\#} \leftarrow partner [i_*, new[i_*]]$ ;
\item~~~
if $j_{\#} = \Lambda$ then $state[i_*, new[i_*]] \leftarrow state1$
\item~~~
else  /" case $j_{\#} \neq \Lambda$ "/
\item~~~~~~
if $j_{\#} > 0$ then  /" state update required "/
\item~~~~~~~~~
if $j_{\#} > N$  then /" one-component interaction "/
\\
$~~~~~~~~~~~~~$
$state[i_*, new[i_*]] \leftarrow jump (state1, j_{\#} -N)$
\item~~~~~~~~~
else \{ /" $1 \leq j_{\#} \leq N$, two-component interaction "/ 
\item~~~~~~~~~~~~~
$state2 \leftarrow 
advance(state[j_{\#}, old[j_{\#}]], 
time[j_{\#}, old[j_{\#}]], time[j_{\#}, new[j_{\#}]])$ ;
\item~~~~~~~~~~~~~
($state[i_*, new[i_*]], state[j_{\#}, new[j_{\#}]]) \leftarrow 
jump(state1, state2)$ ;
\item~~~~~~~~~~~~~
$partner[j_{\#}, new[j_{\#}]] \leftarrow -i_*$ ; 
/" negative partner flags no state update for $j_{\#}$ "/
\\
$~~~~~~~~~~~~$\} ;  /" end two-component interaction clause, 
\\
$~~~~~~~~~~~~$end state update required clause, end $j_{\#} \neq \Lambda$ clause "/
\item~~~~~
$new[i_*] \leftarrow old[i_*]$;   $old[i_*] \leftarrow 3-new[i_*]$ ;
\item~~~~~
{\bf P} $\leftarrow {\min}_{j \in A(i_*)} P_{i_* j} $,
where $A(i_*) = 
\{1 \leq j \leq N, j \neq i_*, time[j,new[j]] \geq P_{i_* j} \}$ ; 
\\
$~~~~~~~~~~$if {\bf P} $< + \infty$ then 
$j_* \leftarrow$ an index which supplies this minimum 
(i.e., {\bf P}) ;
\item~~~~~
{\bf Q} $\leftarrow {\min}_{k \in B} Q_{i_* k} $,
where $B =
\{1 \leq k \leq K\}$ ;
\\
$~~~~~~~~~~$if {\bf Q} $< + \infty$ then 
$k_* \leftarrow$ an index which supplies this minimum 
(i.e., {\bf Q}) ;
\item~~~~~
{\bf R} $\leftarrow \min $\{{\bf P}, {\bf Q}\};  
$time [i_*, new[i_*]] \leftarrow$ {\bf R};
\item~~~~~
if {\bf R} $< + \infty$ then
\item~~~~~~~~
if {\bf Q} $<$ {\bf P} then $partner [i_*, new[i_*]] \leftarrow N+k_*$ 
\item~~~~~~~~
else \{  /" case {\bf Q} $\geq$ {\bf P} "/
\item~~~~~~~~~~~
$time[j_*, new[j_*]] \leftarrow$ {\bf R} ;
\item~~~~~~~~~~~
$m_* \leftarrow partner [j_*, new[j_*]]$ ;
\item~~~~~~~~~~~
$partner [i_*, new[i_*]] \leftarrow j_*$ ; 
$partner [j_*, new[j_*]] \leftarrow i_*$ ;
\item~~~~~~~~~~~
if $m_* \neq \Lambda$ and $m_* \neq i_*$ 
then $partner [m_*, new[m_*]] \leftarrow  \Lambda$ ;
\\
$~~~~~~~~~~~$\} /" end {\bf Q} $\geq$ {\bf P} clause "/
\\
$~~~$\}  /" end while loop "/
\end{enumerate}
}
\end{minipage}}}
\caption{
A version of the simulation algorithm with delayed state update
}
\label{fig:algo2}
\end{figure}
 
The encoding of $partner$ is different in the algorithm in 
Figure~\ref{fig:algo2} comparing
with that in Figure~\ref{fig:algo1}.
In the new version,
\begin{equation}
\label{width}
partner[i, new[i]] = \left\{  \begin{array}{ll}
\Lambda                              & \mbox{for an advancement}\\
\mbox{the index of the partner}      & \mbox{for a two-component interaction}\\
N + \mbox{the index of the obstacle} & \mbox{for a one-component interaction}
                              \end{array}
\right.
\end{equation}
assuming the interaction has not been processed yet.
After the interaction has been processed by one participant $i_*$
but not by the other $j_\#$,
$partner[j_\#, new[j_\#]]$ becomes negative
to indicate that no state update by the second participant
$j_\#$ is required.
This is so done because processing for ball $i_*$ has updated both states.

This code saves not a great deal in the billiard case
because here $advance$ and $jump$ are much lighter computationally
than $interaction\_time$.
The update pattern of array $time[1:N,1:2]$ 
in the algorithm in Figure~\ref{fig:algo2} is the same
as in the one in Figure~\ref{fig:algo1}.

{\bf When the third party is the first party ?}
In both versions of the algorithm,
the third party update 
is conditioned to $m_* \neq i_*$
(lines 18, 19 and 20 in Figure~\ref{fig:algo1}
and line 23 in Figure~\ref{fig:algo2}),
which requires the third party to be distinct from
the first party, the initiator of the update.
The existing program for billiard balls is
supposed to report an occurrence of $m_* = i_*$.
In all the runs,
this condition has never been reported.
Is identity $m_* = i_*$ at all possible ?

We can imagine a scenario when 
equality $m_* = i_*$ is caused by
two components interacting twice,
second interaction immediately after the first one
without intermediate involvement of other components or obstacles.
In the billiard case with periodic boundary conditions
two subsequent collisions of the same pair of balls
is highly improbable for large $N$.
For small $N$, e.g., $N=2$, if balls are relatively large, 
this can occur with a high probability.
In other systems such occurrences may be probable
even for large $N$.
That is why in the general algorithms,
the execution is safeguarded
with the test $m_* \neq i_*$.

\section{Consistency of basic operations}\label{sec:consist}
\hspace*{\parindent} 
In the billiards application, 
the three basic functions of Section~\ref{sec:basic} are defined
in terms of a consistent model:
by integrating differential equations of motion of a system,
using conservation laws, etc.
However, the formulation of the algorithm in Section~\ref{sec:algo} 
employs no additional model.
Clearly, arbitrarily ``bad'' basic functions can cause
arbitrarily bizarre behavior even of a ``good'' algorithm.
Thus, if we wish to provide certain assurance
that the algorithm is ``correct,''
we should request certain ``correctness'' 
or consistency properties
of the basic functions.
Thus, we introduce the following conditions:
\\
\\
(I) Function $interaction\_time$ 
is commutative with respect
to the components,
i.e., it depends on the unordered pair of components,
although in \eqref{inttime} the two participants in the interaction
are represented in a particular order.
\\
\\
(II) Similarly, function $jump$ depends only
on the unordered pair of arguments. 
This means that assignment
$(new\_state2, new\_state1) \leftarrow jump (state2, state1)$
produces the same $new\_state1$, and $new\_state2$
as assignment \eqref{jump}.
\\
\\
(III) Function $advance(state0, time1, time2)$ satisfies a two-parametrical
semigroup property with respect to its second and third argument,
i.e., for any $t_1 \leq t_2 \leq t_3$ we have
$advance (advance(s, t_1, t_2 ), t_2, t_3 ) 
= advance (s, t_1, t_3)$ for any state $s$.
\\
\\
(IV) Moreover, there is a proper associativity 
between $advance$ and $interaction\_time$,
namely, if $t = interaction\_time(s_1, t_1, .)$,
and
$t_1 \leq t_2  \leq t$, then
$t = interaction\_time (advance (s_1, t_1, t_2 ), t_2, .)$,
where dot ($.$) replaces an appropriate pair $(state, time)$
if we have a two-component interaction
or it replaces an $obstacle$ if we have a one-component interaction.
For a two-component $interaction\_time$,
this property, 
coupled with (I), implies a similar associativity with respect
to the second set of arguments or with respect to both sets.
\\
\\
(V) The components are never stuck at each other.
Namely, 
if two components 1 and 2 
with $state1$ and $state2$ are interacting,
i.e. \eqref{already} holds,
$jump$ is applied and $new\_state1$ and $new\_state2$ are computed 
according to \eqref{jump},
then $interaction\_time (new\_state1, time, new\_state2, time) > time$.
Similarly, 
if a component with $state$ is interacting with $obstacle$,
i.e. \eqref{alre0dy} holds,
$jump$ is applied and $new\_state$ is computed according to \eqref{j0ump},
then $interaction\_time (new\_state, time, obstacle ) > time$.
\\

Computationally conditions (I) - (IV)
might be ``slightly'' violated
because of the roundoff.
This can cause dependence
of the simulated history on the processing order.
In double-precision billiard simulations,
a few dozen collisions experienced by each ball is usually sufficient
for two differently ordered computations to completely diverge,
even when started with identical initial conditions.
After the divergence the ball trajectories and the collisions
that occur in one such run are completely different
from the trajectories and collisions that occur in the other.
Computational
physicists are aware of such divergence \cite{ERPW}
and consider it a variant of physical irreproducibility.
It is worth stating however that 
the second run of exactly the same serial program 
starting with the same input data
produces exactly the same results.

Now we are going to introduce a condition of a different kind.
Consider the set of components and obstacles $I(t)$ interacting
at a particular time $t$.
If $I(t)$ is non-empty,
we may introduce a binary relation 
among the elements in $I(t)$,
assuming element $i$ is in the relation with element $j$ 
if $i$ is interacting with $j$ at $t$.
Let $\sim$ be a reflexive, symmetric, and transitive 
closure of this relation, so that $\sim$ is an equivalence.

With this definition, the condition is:
\\
\\
(VI) No equivalence class for the relation $\sim$
contains more than two elements.
\\

\begin{figure}
\centering
\includegraphics*[width=3.2in]{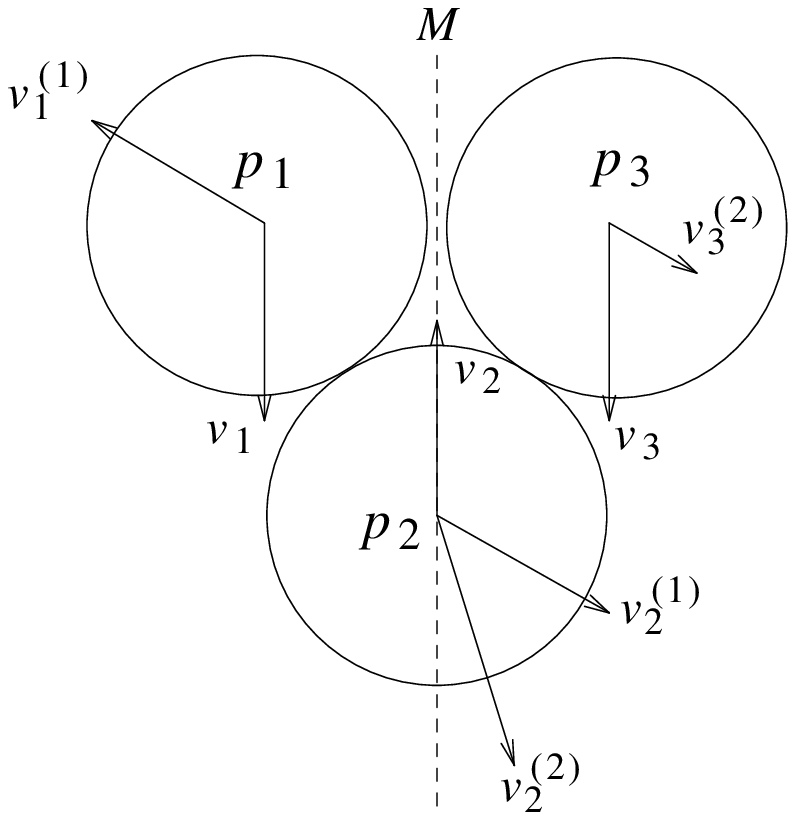}
\caption{A triple collision}
\label{fig:triple}
\end{figure}

In the billiard case, (VI) prohibits, for example,
participation of more than two balls
in the same collision
(but several disjoint pairwise collisions may take place
at the same time).
Figure~\ref{fig:triple} shows such prohibited triple collision
where
\eqref{colliding} holds for the pair $(i=1,j=2)$
and, separately, for the pair $(i=2,j=3)$,
but not for the pair $(i=1,j=3)$,
because $|p_1 -p_3 | > D$.

In Figure~\ref{fig:triple},
the initial condition before collision,
including
positions of the balls  
and their velocities 
$v_1,v_2$  and $v_3$, is 
mirror symmetrical with respect to the middle vertical line $M$.
There are two possible orders 
of processing this collision by the algorithm.
In one order,
balls $1$ and $2$
collide first and
obtain new velocities ${v_1}^{(1)}$ and ${v_2}^{(1)}$,
and then balls
$2$ and $3$ collide
and obtain new velocities ${v_2}^{(2)}$ and ${v_3}^{(2)}$.
The initial velocity of ball $2$ for the second pairwise collision
is ${v_2}^{(1)}$ as if 
the second collision occurred later than the first one.
The net result of the triple collision is the three balls
moving away from the collision site with velocities
${v_1}^{(1)}$, ${v_2}^{(2)}$,
and ${v_3}^{(2)}$,
which are not mirror symmetrical with respect to $M$.
Hence,
the outcome of the triple collision 
depends on the order of processing
and so the history of the entire simulation.

With infinite precision computations,
in the case of chaotically colliding billiard balls,
the probability of violating (VI) is zero.
However, in our finite precision experiments
multiple collisions could practically occur and hence (VI) could be violated.
The proof in Section~\ref{sec:proof} of correctness 
of the simulated trajectory
should be understood as an assurance
that if the machine precision is infinite,
the correctness holds for as long as (VI) holds.

In order to show that the algorithm in Figure~\ref{fig:algo1}
reconstructs the trajectory of each component
``correctly'' we must know what a ``correct''
trajectory is.
With
assumptions (I) - (VI),
starting with a global state at time 0, 
we can uniquely
define the system state at any positive time
using the naive algorithm discussed in Introduction.
We {\em call} the obtained trajectory
the correct one.

The algorithm in Figure~\ref{fig:algo1} 
ignores many events on this trajectory.
The task of proof is showing that despite of this
the trajectory does not change.

\section{Invariants and correctness proof}\label{sec:proof}
\hspace*{\parindent} 
The actions of either simulation algorithms in Figure~\ref{fig:algo1}
or in Figure~\ref{fig:algo2} can be summarized as follows:
a repeated update of arrays $new[1:N], old[1:N]$, and $event[1:N,1:2]$ 
in such a way that the following conditions 
remain invariant:
\begin{equation}
\label{invar1}
\max_{1\leq i\leq N} time [i, old[i]] \leq \min_{1\leq i\leq N} time [i,~new[i]]
\end{equation}
for $i = 1,2,...N$ we have:
\begin{equation}
\label{invar2}
time[i,new[i]] \leq \min 
\{ \min_{1\leq j\leq N, j \neq i, time[j,new[j]] \geq P_{ij}} P_{ij} ,
\min_{1\leq k\leq K} Q_{ik} \}
\end{equation}
and
\begin{equation}
\label{invar3}
\begin{array}{l}
\mbox{either $partner[i,new[i]] = \Lambda$},\\
\mbox{or $j = partner[i,new[i]]$ 
is an integer in the interval $N+1<j\leq N+K$},\\
\mbox{or $j$ is an integer in the interval $1 \leq j \leq N$ 
and $partner[j,new[j]] = i$}
\end{array}
\end{equation}

Conditions \eqref{invar1}, \eqref{invar2}, and \eqref{invar3} are trivially
satisfied in the beginning of the simulation.
Invariance of condition \eqref{invar1} is obvious.
As to \eqref{invar2} and \eqref{invar3}, 
their invariance can be violated temporarily 
after a cycle during which one component
participating in a two-component interaction
has been processed but the other has not been yet.
After both components have been processed,
and no other two-component interaction processing
has been started,
\eqref{invar2} and \eqref{invar3} hold.
For \eqref{invar2},
it follows from lines 4, 5, and 6 in Figure~\ref{fig:algo1}
and for \eqref{invar3}, it follows
from symmetricity of matrix $P_{ij}$.
The symmetry is an obvious implication of (I) and (II).
Observe, that invariance of \eqref{invar1} and \eqref{invar2} 
requires no consistency conditions (I) - (VI).

\begin{figure}
\centering
\includegraphics*[width=5.8in]{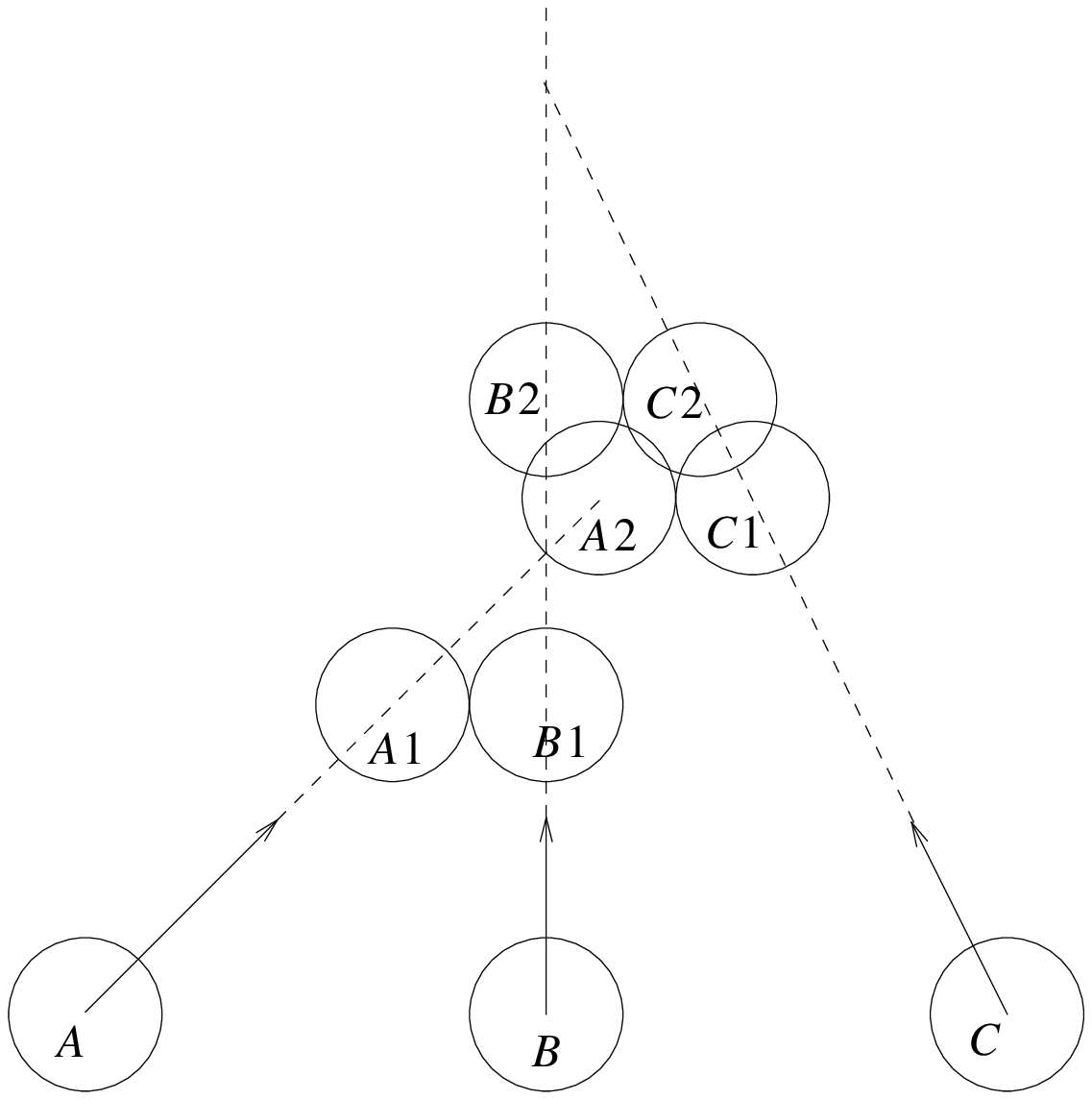}
\caption{Asynchronous collisions of three billiard balls}
\label{fig:asynch}
\end{figure}

Invariant \eqref{invar2}
is the key to understanding the ``wasteful'' strategy
of the data update in this algorithm.
Consider an example.
Let $N = 3, K = 0$.
Figure~\ref{fig:asynch} shows
trajectories of three
billiard balls $A$, $B$, and $C$.
We assume that
at time $t = 0$ the balls are positioned on the same
horizontal line and we suppose that
these are their $old$ positions,
i.e., those stored in array $event[.,old[.]]$.

On the basis of the $old$ events only,
$C$ can see two immediate collisions, one with $B$
when the balls occupy positions $B2$ and $C2$;
we call it collision $B2$,$C2$,
and the other with $A$, namely collision $A2$,$C1$.
$C$ also notes that both $A$ and $B$ have a scheduled
event 
at time earlier than times of either $A2, C1$ and $B2, C2$.
Thus, the set of balls $X$ over which the minimum of $P_{CX}$
is to be taken according to \eqref{invar2} is empty
and this minimum together with the time of the immediate next
interaction for $C$ is $+ \infty$.

With the given $old$ events,
the following assignment of $new$ times would satisfy \eqref{invar2}:
both
$time[A,new[A]]$ and $time[B,new[B]]$ are equal
to the time of collision $A1,B1$,
end $time[C,new[C]] = + \infty$.
With such an assignment,
three inequalities \eqref{invar2} turn into equalities.

The assignment $time[i,[new[i]] = + \infty$
simply means that
$C$ sees no future interaction
at this stage of simulation.
A more aggressive strategy of precomputation,
in which $C$ would look one more step ahead
and would examine
possible collisions with $A$ and $B$
based on their velocities after an as yet 
unprocessed collision $A1,B1$,
is possible but
is prohibited in the proposed algorithm.
Such  an aggressive strategy perhaps would work 
well for a small number of balls
but if there are
many of them,
it would require
a complicated data structure to support
an arbitrary-many-step lookahead.
Following our ``lazy'' or
``wasteful'' strategy, 
$C$ does not do this
and does not schedule
for more than one step ahead.
This allows us to
keep data structure very simple.

Invariants \eqref{invar1} and \eqref{invar2}
imply the following useful invariant
\begin{equation}
\label{invar4}
\min \{ \min_{1\leq i,j \leq  N} P_{ij}, 
        \min_{1 \leq i\leq N, 1\leq k\leq K} Q_{ik} \}  
\geq  \min_{1\leq i\leq N} time[i,new[i]]
\end{equation}

Now we prove correctness of the algorithm.
The proof proceeds by showing that if
\\
\\
(*) the simulated trajectory
is identical to the ``correct'' one defined
in Section~\ref{sec:consist},
for all $t$ in the interval 
$0 \leq t \leq \max_{1\leq i\leq N} time [i, old[i]]$,
\\
\\
then 
\\
\\
(**) after all events with times equal to
$\min_{1\leq i\leq N} time [i, new[i]]$ will be processed,
the thus extended simulated trajectory 
will be identical to the ``correct'' trajectory
for all $t$ in the interval
$0 \leq t \leq \min_{1\leq i\leq N} time [i, new[i]]$.
\\

This would constitute the inductive step,
while the base for the induction is obviously satisfied
since (*) is correct for the program state initialized 
for $t = 0$ as described in Figure~\ref{fig:algo1}.

The ``correct'' trajectory has no interaction on the open interval
\[\max_{1\leq i\leq N} time[i,old[i]] < t < 
\min_{1\leq i\leq N} time[i,new[i]] ,\]
because if it did, \eqref{invar4} would be violated.
Hence the simulated trajectory is identical
to the ``correct'' one for all $t$ in the interval
$0 \leq t < \min_{\leq i\leq N} time[i,new[i]]$.
By (I) - (VI), this property extends to the point 
$t =  \min_{1\leq i\leq N} time[i,new[i]]$
and this completes the proof.

\section{An application example: disk packing problem}\label{sec:pack}
\hspace*{\parindent} 
The following model is simulated in \cite{LSTIL}:
$N$ points are placed randomly 
within an $L \times L$ square.
Periodic boundary conditions apply in both directions.
The $N$ points are assigned random initial velocities 
and in the absence of subsequent collisions would move
with these velocities along straight lines threading through
an infinite sequence of periodic images of the basic square.
However, the points also begin to grow at a common rate into
elastic rigid disks, with diameters that are given by linear
function of time $D(t) = at, t>0$.
As a result, particle collisions become
possible, and increase in frequency as $D(t)$ increases.
We permit $D(t)$ to grow until the system ``jams up''
thus obtaining the final packing.

This is a variant of the billiard simulations.
Two differences are:
\\
1) instead of equation $|p+vt|^2  =  D^2$
as in Figure~\ref{fig:geomean},
equation $ | p + v t |^2  = ( a t )^2$
has to be solved; the latter
is still a quadratic equation with respect to $t$
\\
2) the normal components of 
${v_1}^{new}$ and
${v_2}^{new}$,
the velocities of disks after a collision
(see Figure~\ref{fig:collis}),
have to be increased
to guarantee that disks do not overlap or stick at each other.
Any additive velocity larger than $a/2$ would be appropriate.

Energy or momentum conservation
are lost with such an additive;
as the simulation progresses the system ``heats up''
and as the disk speeds reach large values
computational precision may be lost.
The existing program once in a while interrupts the simulation
projecting all the disk positions
into a particular time value,
then scales down and balances them.
(The velocities $v_i , i=1,...N$,
of $N$ disks of equal masses
are said to be balanced if $\sum_{1\leq i\leq N} v_i = 0$.)

\begin{figure}
\centering
\includegraphics*[width=5.8in]{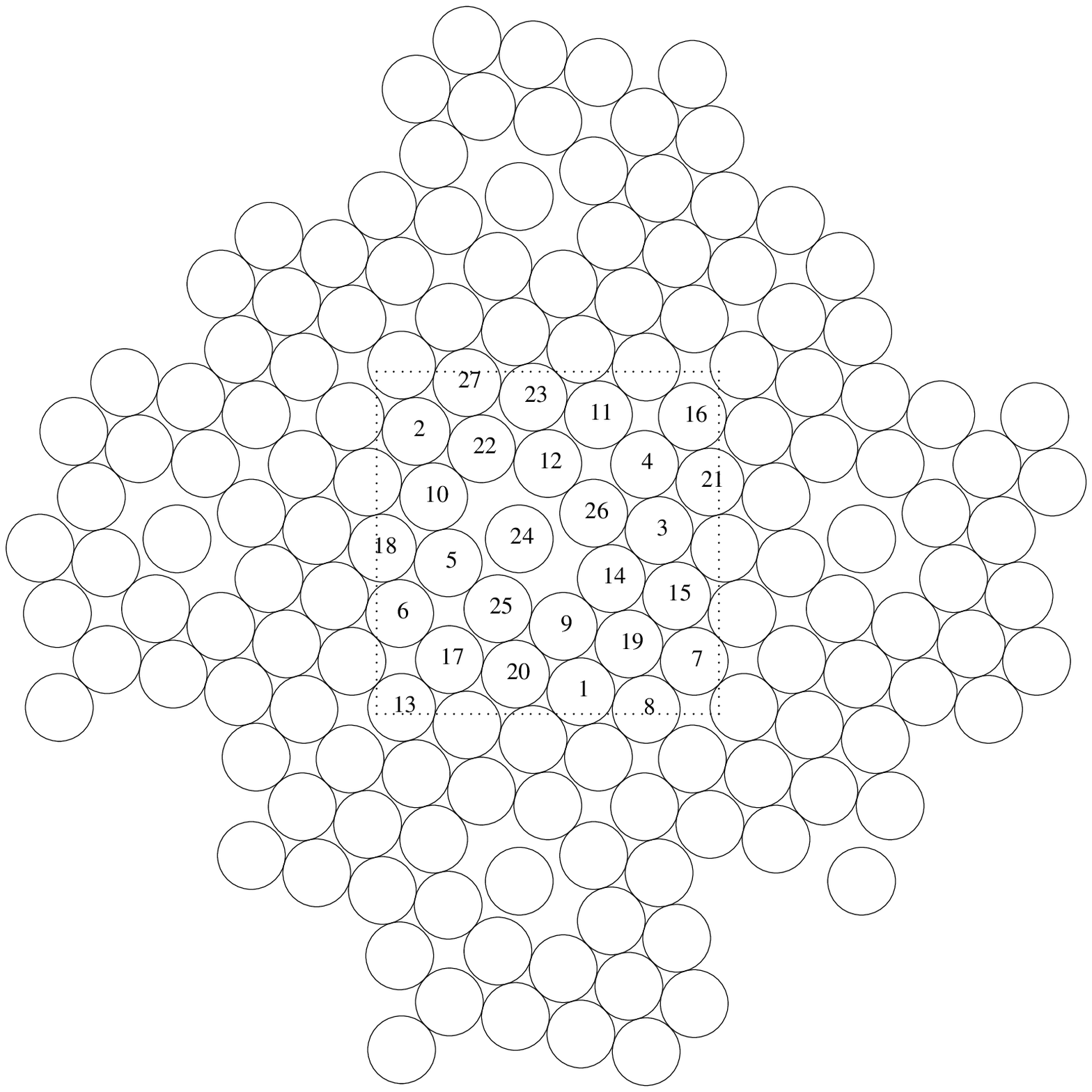}
\caption{27 disks after 100000 collisions; disk 24 is a rattler}
\label{fig:27balls}
\end{figure}

\begin{figure}
\centering
\includegraphics*[width=5.8in]{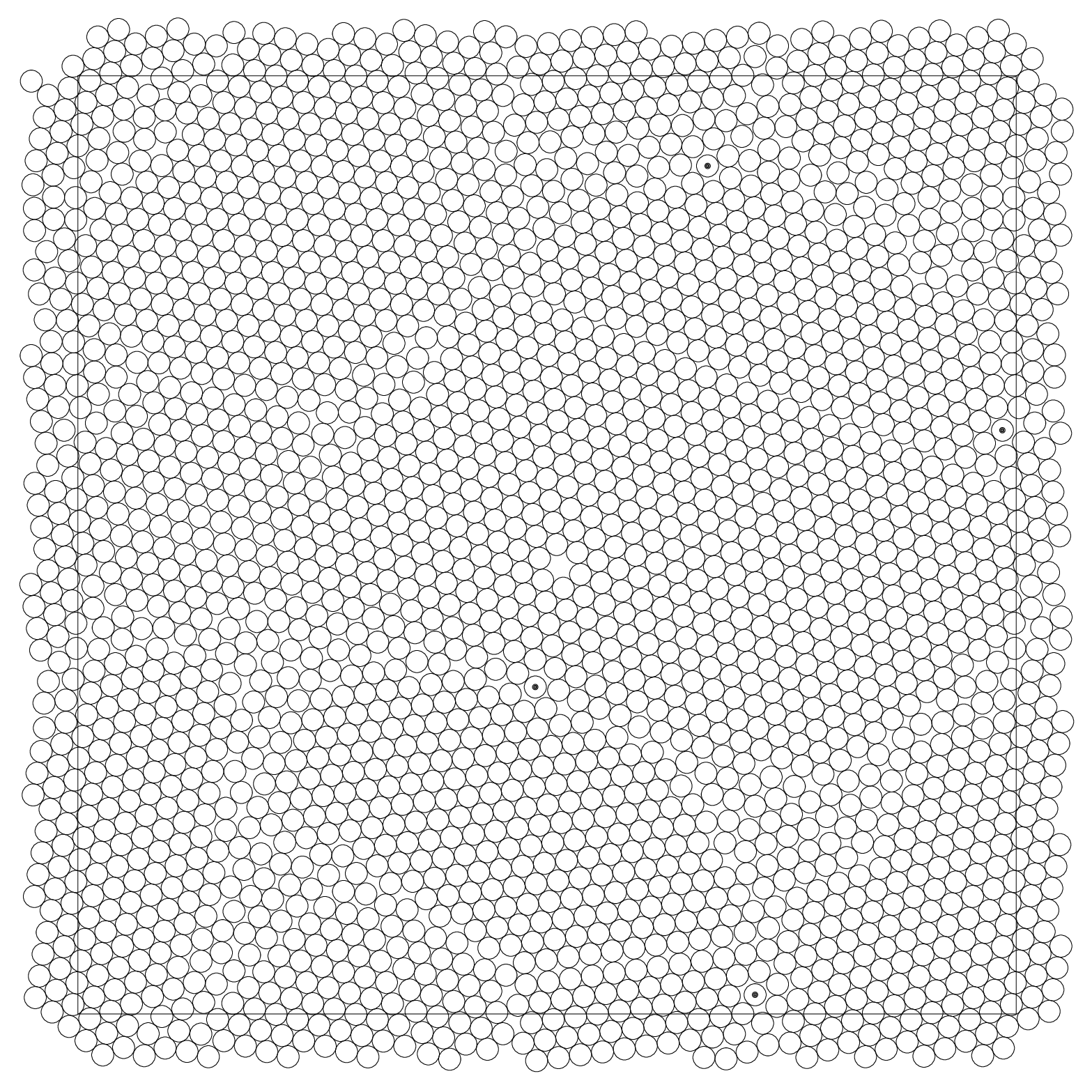}
\caption{2000 disks after $42 \times 10^6$ collisions. 
Dots mark significant rattlers}
\label{fig:2000ball}
\end{figure}

Figure~\ref{fig:27balls} and \ref{fig:2000ball} 
show some results of these experiments,
in particular show ``rattler'' disks which remains unjammed
within the walls of jammed neighbors \cite{LSTIL}.
In the experiment presented in Figure~\ref{fig:2000ball},
the large square is subdivided into $40 \times 40$
small square sectors.
Rather than checking a possible next
collision with $8 \times 1999$ candidates, only about 
10 disk candidates for the next collision are checked.

\section{Comparing practical performance of the algorithm with the other published algorithms}\label{sec:perf}
\hspace*{\parindent} 
Physicists often study
hard-sphere and hard-disk models using computing experiments.
However, with the exception of \cite{ERPW} and \cite{ALDER},
nobody discusses the details of the algorithms used
and
with the exception of 
\cite{ALDER},
nobody gives performance data.

We read in \cite{ALDER}:
``The IBM704 calculator handles about
2000 collisions per hour for 100 molecules
and about 500 collisions per hour for 500 molecules''

Assuming that IBM704 was not slower than 0.02 MFLOP
\cite{BASH} this scales up to no more than 30 collisions per second for
a 1 MFLOP machine.
The speed of our calculations
is in the range 150 - 450 pairwise collisions
per second (independently of the number of disks)
on VAX8550 which has speed 1 MFLOP.
Thus, even the most pessimistic comparison with \cite{ALDER}
gives about an order of magnitude speed-up of our algorithm.

Simulation of 50000 to 55000
committed events in a random configuration of 160 disks
is reported in \cite{HONT}.
Let us count one pairwise collision 
as two committed events,
and one sector boundary crossing 
or cushion reflection as one committed event.

The model \cite{HONT} is different from the one we simulated in that
instead of periodic boundary conditions,
rigid elastic ``cushions'' are employed to guard the cell boundaries.
To compensate for the difference,
let us equate an external boundary
crossing in program \cite{LSTIL}
with one cushion reflection in \cite{HONT}.
Note that when scheduling a collision 
close to the cell boundary,
program \cite{LSTIL} 
considers not only internal disks as the candidates
for collision,
as program \cite{HONT} does,
but also their periodic images.
This additional complexity in program \cite{LSTIL} 
more than 
compensates for a possible loss of complexity
due to substituting 
a cushion reflection with a boundary crossing.

To make a fair comparison,
ratio of the diameter to $L$ must be set as in \cite{HONT}.
Since the value of this ratio is not indicated in \cite{HONT},
different pictures of 160 randomly placed disks
for several different diameters were produced
and the one which resembled the most
Figure 9 in \cite{HONT} was selected.
In the selected picture, the ratio is 0.015.
Then we made a measuring run,
in which
sector boundary crossings
were counted only for 16 sectors as specified
in series I in \cite{HONT}.
The run was continued
until the number reached 52000 as in \cite{HONT}.
It took
90 seconds CPU to complete this run.

A similar run in \cite{HONT} (Series I) 
on one PE took 440 seconds
and took 62 seconds
on 32 nodes of a hypercube MARK III.
(For 32 and 64 sectors, it took, respectively, 
44 and 42 seconds in \cite{HONT}.)
It is known \cite{FOX} that one node of MARK III is about 60\% 
faster than a VAX 8550, the host computer in \cite{LSTIL}.
Besides, algorithm \cite{LSTIL} is a Fortran code 
while program \cite{HONT} is written in the C-language,
both compiled under $UNIX^{TM}$. 
This
yields an additional 10\% in favor of algorithm \cite{LSTIL},
since Fortran is slower than C under a $UNIX^{TM}$ compilation.

Thus, one concludes that the serial algorithm \cite{LSTIL} runs about
as fast as the parallel Time Warp \cite{HONT} on a 32-node hypercube.
\footnote{
After the measuring run was completed,
A.P. Wieland kindly informed the author that
one sector boundary crossing is actually counted 
as two events in \cite{HONT}
rather than as one, as is
assumed in this paper.
Also, one pairwise disk collision is counted
not as two events as assumed,
but as $4+m$ events, where variable $m$
is the number of disks located in the involved
sectors at the time of the collision.
Suppose only one sector is involved in each collision,
and there are totally 16 sectors 
(as in Series I in \cite{HONT}).
Then $m$ is about 10.
This makes the total count of events generated in \cite{HONT}
during a comparable simulation time interval several times
higher than was assumed in the experiments.
This, in turn, makes program
\cite{LSTIL} faster than program \cite{HONT}.
}

\section{Other billiard-like simulations}\label{sec:other}
\hspace*{\parindent} 
A collision of two billiard balls of radius $D/2$ each
can be considered as an interaction of two zero-size particle
with potential $V(r) = 0$ for distances $r > D$ and 
$V(r) = + \infty$ for $r \leq D$.
More general piece-wise constant potentials
can be dealt with using the same algorithm,
e.g.,
the square-well potential \cite{ALDER}:
\begin{equation}
\begin{array}{ll}
V = + \infty , & \mbox{if}  r <  \sigma_1\\
V = V_0 ,      & \mbox{if} \sigma_1 < r <  \sigma_2\\
V =  0 ,       & \mbox{if} \sigma_2 < r
\end{array}
\end{equation}
where $V_0$, $\sigma_1$, and $\sigma_2$ are finite constants.
We can imagine two concentric balls: the ``hard core'' ball
of diameter $\sigma_1$ and a larger external ball of diameter $\sigma_2$
and correspondingly two types of ``collisions'':
internal, of the hard-cores and of the external balls.
Each type has its own $jump$ function.

By the Monte-Carlo simulation 
\cite{ROSATO} shows that larger particles move against
the gravitational force if placed in a vibrating container together
with smaller particles.
The balls of different diameters can be easily handled in our scheme,
if the ball diameter becomes a part of its $state$.
The model \cite{ROSATO}
can be easily simulated using direct representations
of particle dynamics, rather than by Monte-Carlo.

The inhomogeneity of components,
in general,
can be treated in the same way, i.e., by making
the type or the class identification of a component
an unchangeable part of its $state$.
Combat simulations \cite{WIEL} present 
such inhomogeneity to a large extend,
since here a component represents a military unit
of one of the two opposing armies
and there are many types of such units.

Collisions may be generalized to any state
changes, not necessarily immediately leading to the
trajectory change.
A typical simulation rule in \cite{WIEL}
looks like follows:
``if within radius $\sigma$, a unit detects $m$ units of the same army 
and $n$ units of the opposing army,
then it takes course of action $c(n,m)$
from the time of detecting this situation until 
the time when
another rule becomes applicable.''
We can represent the set of these rules
by surrounding a zero-sized unit by several
circles, each representing one such a rule,
A counter ``inside'' the unit state gets an instantaneous increment,
when a particular circle ``collides'' with another unit.
The counter change may or may not 
trigger the change of the course of the actions.
Such mechanisms can be represented within the discussed
framework and simulated using the algorithm in Figure~\ref{fig:algo1}.

According to \cite{SMITH},
a variant of the dense packing algorithm
can be used in finding optimal spherical codes.
Here the task is to find $N$ 
points $p_i ,i = 1,...N$, on the sphere in the $k$-dimensional
Euclidean space in such a way that
$\min_{i\neq j} \mbox{distance}(p_i, p_j) \rightarrow \max$.
We would start with a random configuration of
$N$ points and then grow ``caps''
of equal size each cap having one point as the center.
Caps are prevented from the overlap by the mean of collisions.


\end{document}